\documentclass[sigconf]{acmart}

\usepackage{enumitem}
\usepackage[utf8]{inputenc}
\usepackage{multirow}
\usepackage{float}
\usepackage{subcaption}
\copyrightyear{2026}
\acmYear{2026}
\setcopyright{cc}
\setcctype{by-nc-nd}
\acmConference[CHI '26]{Proceedings of the 2026 CHI Conference on Human Factors in Computing Systems}{April 13--17, 2026}{Barcelona, Spain}
\acmBooktitle{Proceedings of the 2026 CHI Conference on Human Factors in Computing Systems (CHI '26), April 13--17, 2026, Barcelona, Spain}
\acmPrice{}
\acmDOI{10.1145/3772318.3791248}
\acmISBN{979-8-4007-2278-3/2026/04}

\begin{document}

\title{InnerPond: Fostering Inter-Self Dialogue with a Multi-Agent Approach for Introspection}

\author{Hayeon Jeon}
\orcid{0009-0003-5864-1185}
\affiliation{
  \institution{hci+d lab.}
  \institution{Seoul National University}
  \city{Seoul}
  \country{Republic of Korea}
}
\email{jhy94520@snu.ac.kr}

\author{Dakyeom Ahn}
\orcid{0009-0004-9139-1605}
\authornote{These authors contributed equally to this research.}
\affiliation{
  \institution{hci+d lab.}
  \institution{Seoul National University}
  \city{Seoul}
  \country{Republic of Korea}
}
\email{adklys@snu.ac.kr}

\author{Sunyu Pang}
\orcid{0009-0002-5486-2047}
\authornotemark[1]
\affiliation{
  \institution{hci+d lab.}
  \institution{Seoul National University}
  \city{Seoul}
  \country{Republic of Korea}
}
\email{sunyu.pang@snu.ac.kr}

\author{Yunseo Choi}
\orcid{0009-0008-6929-404X}
\authornotemark[1]
\affiliation{
  \institution{hci+d lab.}
  \institution{Seoul National University}
  \city{Seoul}
  \country{Republic of Korea}
}
\email{dbstj0531@snu.ac.kr}

\author{Suhwoo Yoon}
\orcid{0009-0004-5893-5439}
\affiliation{
  \institution{hci+d lab.}
  \institution{Seoul National University}
  \city{Seoul}
  \country{Republic of Korea}
}
\email{yeopil@snu.ac.kr}

\author{Joonhwan Lee}
\orcid{0000-0002-3115-4024}
\affiliation{
  \institution{hci+d lab.}
  \institution{Seoul National University}
  \city{Seoul}
  \country{Republic of Korea}
}
\email{joonhwan@snu.ac.kr}

\author{Eun-mee Kim}
\orcid{0000-0003-4032-4731}
\authornote{Co-corresponding authors}
\affiliation{
  \institution{Department of Communication}
  \institution{Seoul National University}
  \city{Seoul}
  \country{Republic of Korea}
}
\email{eunmee@snu.ac.kr}

\author{Hajin Lim}
\orcid{0000-0002-4746-2144}
\authornotemark[2]
\affiliation{
  \institution{hci+d lab.}
  \institution{Seoul National University}
  \city{Seoul}
  \country{Republic of Korea}
}
\email{hajin@snu.ac.kr}

\renewcommand{\shortauthors}{Hayeon Jeon et al.}

\begin{abstract}
Introspection is central to identity construction and future planning, yet most digital tools approach the self as a unified entity. In contrast, Dialogical Self Theory (DST) views the self as composed of multiple internal perspectives, such as values, concerns, and aspirations, that can come into tension or dialogue with one another. Building on this view, we designed InnerPond, a research probe in the form of a multi-agent system that represents these internal perspectives as distinct LLM-based agents for introspection. Its design was shaped through iterative explorations of spatial metaphors, interaction scaffolding, and conversational orchestration, culminating in a shared spatial environment for organizing and relating multiple inner perspectives. In a user study with 17 young adults navigating career choices, participants engaged with the probe by co-creating inner voices with AI, composing relational inner landscapes, and orchestrating dialogue as observers and mediators, offering insight into how such systems could support introspection. Overall, this work offers design implications for AI-supported introspection tools that enable exploration of the self's multiplicity.
\end{abstract}



\maketitle

\section{Introduction}

\begin{figure*}[t]
    \centering
  \includegraphics[width=0.75\textwidth]{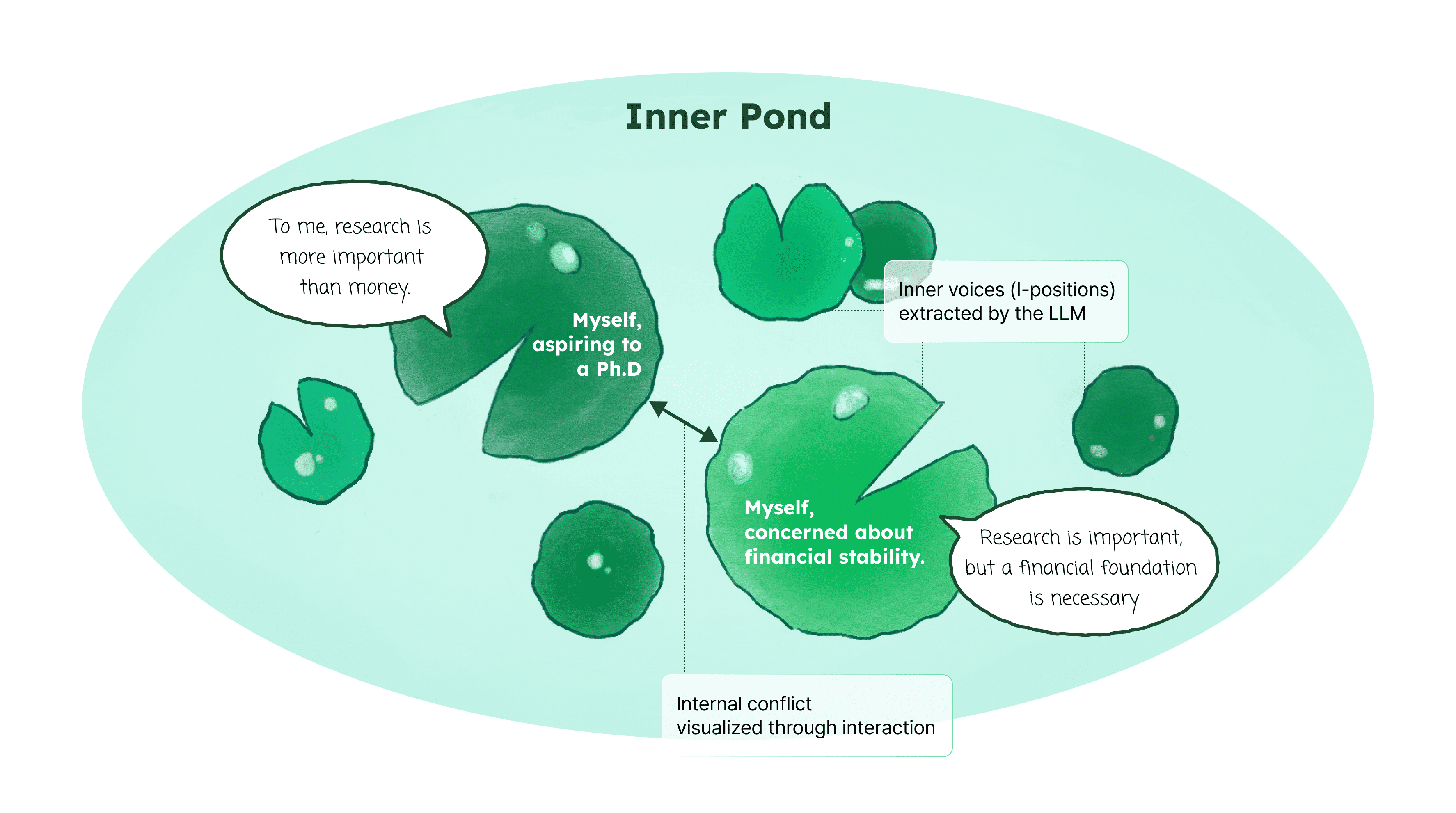}
  \caption{InnerPond fosters AI-mediated introspection through multi-agent dialogue. The system transforms users' I-positions into conversing LLM agents (lotus leaves), supporting dialogue among multiple I-positions within the self.
}
  \Description{A conceptual visualization of the InnerPond system where users' I-positions become conversing LLM agents represented as lotus leaves on a pond. The image illustrates two example I-positions with contrasting perspectives on career priorities engaging in dialogue. Additional lotus leaves represent other I-positions, demonstrating the system's multi-agent approach to supporting dialogue among multiple I-positions within the self.}
  \label{fig1}
\end{figure*}

Individuals constantly face complex life choices \cite{savioni2023make}, where introspection---the critical examination of one's own thoughts, values, and emotions \cite{van2000inward}---plays a crucial role in navigating these decisions \cite{boring1953history,byrne2005introspection}. A key mechanism that facilitates introspection is inner dialogue, in which different internal perspectives engage in a mental conversation \cite{fernyhough2023inner, oles2020types, puchalska2015self}. To better understand the complexities of this inner dialogue, the \textbf{Dialogical Self Theory (DST)} \cite{hermans2011handbook, hermans1996voicing} provides a theoretical framework. Rather than viewing the self as an isolated, unified entity, DST conceives it as a relational process---a ``dynamic multiplicity of I-positions'' \cite{hermans2001dialogical}. Put simply, our mind hosts various `selves'---like `the creative self' or `the fearful self'---each possessing a distinct perspective and narrative \cite{hermans2014self}. These I-positions are not isolated; they engage in a constant dialogue, influencing and repositioning one another within a dialogical space \cite{hermans2003construction, hermans2010m}. Through such interaction, a more reflective and emergent self-understanding could emerge \cite{hermans2011handbook, hermans2014self}.

While inner dialogue is a vital mechanism for introspection, in its spontaneous form, it may remain constrained within familiar perspectives or spiral into unproductive rumination \cite{boudreault2018investigation, fritsch2024relationship}. To scaffold this process more constructively, various aids for introspection have been explored---from artifacts such as journals, photos, and music \cite{staahl2008reflecting, chen2019chronoscope, petrelli2014photo} to more recent LLM-based systems \cite{jeon2025letters, song2025exploreself}. These approaches have provided valuable ways to foster introspection. However, by framing users primarily as singular, unified selves, they have given less attention to the plural and dynamic nature of the self---particularly the way multiple internal voices coexist, compete, and negotiate with one another during complex life choices \cite{hermans2011handbook, hermans2014self}.

We aim to address this gap by translating DST's concept of a dialogical self into an interactive system and, through this instantiation, generating design knowledge about supporting dialogue among one's multiple selves. In particular, LLMs can now embody distinct personas with consistent perspectives \cite{cheong2022role, shao2023character}, enabling new ways to externalize this plurality. Leveraging this capability, we investigate a design exploration that makes inner multiplicity explicit---representing multiple I-positions as distinct AI agents that can dialogue with one another and with the user. Through this approach, we explore how people experience and make sense of such multi-agent inner dialogue. Building on this, we address the following research questions (RQs):
\begin{itemize}
    \item \textbf{RQ1:} What are the key design considerations for supporting dialogue among multiple selves in exploring the plural and dynamic self?
    \item \textbf{RQ2:} How do people experience and make sense of engaging with their multiple selves through multi-agent AI-mediated introspection?
\end{itemize}

In addressing the RQs, we designed and developed \textbf{InnerPond} as a research probe \cite{hutchinson2003technology, boehner2007hci, ccercci2021design}---not to evaluate system effectiveness, but to inspire design exploration of how people experience dialogical engagement with their multiple I-positions. Informed by design considerations around spatial metaphors, scaffolding structures, and multi-agent conversational dynamics, InnerPond represents an individual's various I-positions as independent LLM-based agents that can be visualized, elaborated, and brought into dialogue within a shared space. Anchored in the lotus leaf and pond metaphor (Figure \ref{fig1}), InnerPond aims to enact coexistence and interconnection among multiple selves, helping users explore their inner landscape and move toward more emergent forms of self-understanding.

To understand how people experience such engagement with their multiple selves, we conducted a user study (N=17) with young adults deliberating between two career paths---a context where competing internal voices often surface \cite{amir2006facets, taber2013time} and structured support can help navigate their complexity \cite{kocielnik2018reflection, bentvelzen2022revisiting}. Our findings showed that through externalizing and dialoguing with their inner voices, participants could develop a meta-cognitive perspective---recognizing distinct selves as interconnected parts of a larger self. For some, this shift reframed how they approached career decisions, moving from reacting to external pressures toward reflecting on internal alignment.

Through this exploration, we use the term ``\textit{inter-self communication}'' to describe dialogue among one's own inner voices, as conceptualized in DST. Unlike solitary reflection, it gives distinct voices space to interact, making the multiplicity of the self more tangible; unlike dialogue with others, it keeps exchanges grounded within the self. This perspective opens broader possibilities for designing AI systems that support people in exploring, elaborating, and engaging with their plural selves. We discuss the implications of this approach, considering both the opportunities it offers and the design tensions that accompany it.

\begin{figure*}[!ht]
\centering
\includegraphics[width=0.6\textwidth]{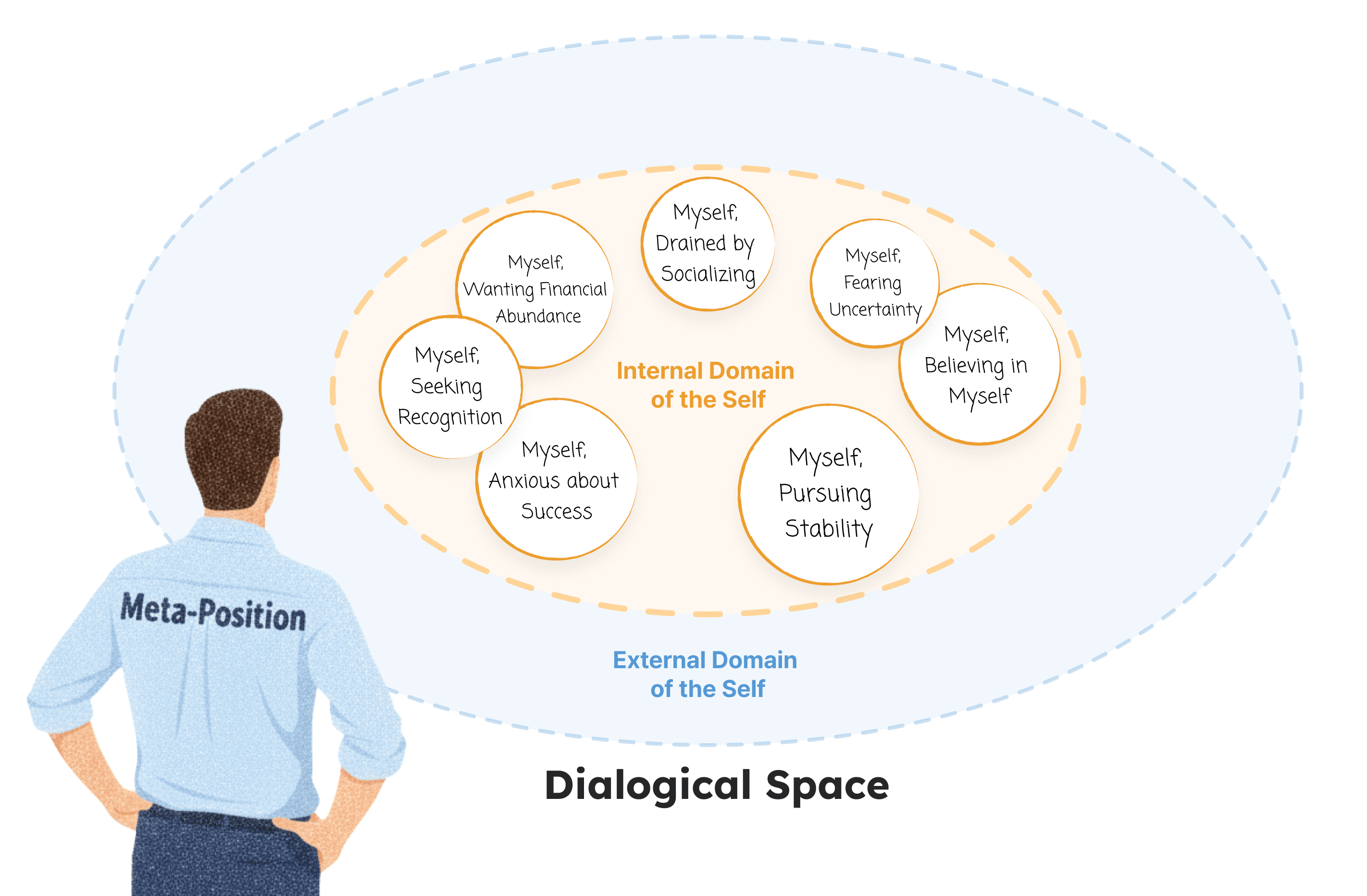}
\caption{Core concepts of Dialogical Self Theory, illustrating the relationship between multiple I-positions in the dialogical space and a meta-position.}
\Description{This figure illustrates the core concepts of Dialogical Self Theory. The internal domain of the self contains multiple I-positions (e.g., `Myself, Seeking Recognition,' `Myself, Pursuing Stability'), each representing distinct perspectives within the society of mind. The meta-position, depicted as an observer, provides a higher-level vantage point for reflecting on and integrating these diverse internal voices. The dialogical space encompasses both internal and external domains where these I-positions engage in dynamic interactions.}
\end{figure*}

Taken together, this paper makes two contributions to HCI:
\begin{itemize}
    \item \textbf{Design considerations for plural and dynamic self engagement} --- Informed by our design journey with InnerPond, we articulate key considerations for translating DST into interactive experience: spatial metaphors that convey coexistence and interconnection, scaffolded stages that structure the introspective journey, and multi-agent dynamics that balance consistency with flexibility. These considerations extend current approaches to AI-supported introspection by shifting from a unified to a dialogical view of the self.
     \item \textbf{Empirical insights into inter-self dialogue} --- We provide rich qualitative insights into how people experience and make sense of engaging with their multiple selves, revealing how inter-self dialogue is experienced and interpreted when instantiated through a multi-agent system. Our findings surface both opportunities and design tensions, offering implications for future systems that aim to support the plural and dynamic nature of the self.
\end{itemize}

\section{Related Work}

This section examines inner dialogue as a core mechanism of introspection and introduces Dialogical Self Theory. We then review existing approaches to supporting introspection. Finally, we highlight the potential of multi-agent systems as a means to support dialogical forms of introspection.

\subsection{Dialogical Self Theory as Framework for Understanding Inner Dialogue}
Introspection is a practice of looking inward and exploring one's own thoughts, values, and emotions \cite{byrne2005introspection}. In psychology, it is characterized as explicitly observing and reflecting on an individual's mental state \cite{boring1953history}. Distinguished from simply reflecting on or recollecting memories, introspection involves critically evaluating significant experiences and patterns in one's life, and continuously questioning what has been achieved and what is desired \cite{van2000inward}. Introspection and self-reflection are closely related---both involve examining one's mental states to foster self-understanding \cite{frank2019mindfulness, arnera2024digital}. However, they differ in scope: self-reflection refers to the cognitive process of examining one's thoughts and actions, whereas introspection encompasses a broader philosophical construct that includes emotional and evaluative dimensions \cite{hoyer2000self}. Through this process, individuals search for answers concerning their identity, values, and life priorities, fostering deeper self-understanding and identity construction \cite{brand2023envisioning, brand2021design}. Particularly in life choice contexts such as career exploration, active introspection on one's identity and life directions plays a crucial role \cite{ran2023linking}.

A key internal process facilitating introspection is \textit{inner dialogue}  \cite{fernyhough2023inner}. Inner dialogue is a process where ``different internal perspectives engage in mental conversation,'' exchanging views and positions within the mind \cite{oles2020types, puchalska2015self}. Unlike simple monologues, inner dialogue reflects the dialogical nature of the self and simulates social dialogical relations internally \cite{alderson2018varieties, hofman2016history}. Through confronting and integrating diverse internal perspectives, it enables reflexive and contemplative self-understanding and identity construction \cite{puchalska2020functions, latinjak2023self}.

To better conceptualize this process, \textit{Dialogical Self Theory (DST)} \cite{hermans2011handbook, hermans1996voicing} offers a structured framework for understanding the complex dynamics of inner dialogue. It conceptualizes inner dialogue as an exchange of thoughts and perspectives between distinct I-positions within the self, helping identify and interpret various forms of internal dialogical activity during introspective processes \cite{oles2020types}. DST reconceptualizes the self not as an isolated inner entity but as an inherently relational process---a dialogical being described as a dynamic multiplicity of I-positions within the society of mind \cite{hermans2001dialogical}. Each I-position represents a narratively structured unit of the self with its own perspective and voice. For example, I-positions such as ``I as dreamer'' or ``I as fearful'' reflect different internal aspects of the person \cite{hermans2014self}, engaging in dialogical relationships \cite{hermans2003construction}.

These I-positions are shaped as external voices become internalized within the self. Through multidimensional inner dialogue unfolding in a dialogical space, these I-positions are dynamically repositioned through DST's core mechanisms---positioning, counter-positioning, and repositioning---with their relationships continually restructured \cite{hermans2011handbook, hermans2010m}. DST further introduces the concept of the `\textit{meta-position}', a higher-level vantage point from which a person can observe and reflect on their various I-positions. This capacity for meta-level reflection serves as a crucial mechanism for self-understanding and integration \cite{hermans2011handbook, hermans2003construction, hermans2014self}.

This view resonates with recent theorizing in personal informatics, where the self has been reconceptualized as dynamic and constructed through ongoing interaction with the world \cite{rapp2017know}. Yet, while such work emphasizes the self's temporal and social dimensions---how it evolves across past, present, and future, and in various social contexts---it gives less attention to the dialogical mechanisms through which distinct aspects of the self actively engage with one another. DST complements this perspective by foregrounding these inter-positional dynamics, offering a foundation for understanding introspection not merely as self-observation, but as a process through which self-understanding emerges from dialogue among multiple internal voices.

\subsection{Existing Approaches to Supporting Introspection}
While inner dialogue serves as a key mechanism for introspection, its naturally occurring forms are often limited by the individual's existing experiences and perspectives, resulting in constrained self-understanding \cite{boudreault2018investigation}. In situations where thoughts and emotions are deeply intertwined, individuals can fall back on habitual thinking patterns or become trapped in negative cycles such as rumination \cite{fritsch2024relationship}. To address these challenges, various introspective interventions have been developed.

Traditional introspective interventions have supported self-reflection through personal artifacts such as photographs, letters, music, and diary writing \cite{belk1988possessions, belk1990role, kleine1995possession, shoemaker1986introspection}. Among these, diary writing has been one of the most common methods for critically evaluating past experiences and questioning the future \cite{o2005diaries}. However, these writing-based approaches, grounded in a person's existing perspective \cite{mols2012dear}, may have limitations in offering new insights or alternative viewpoints \cite{mols2016informing}.  

Further, digital technologies have expanded these practices, incorporating journals \cite{staahl2008reflecting, park2025nore}, photographs \cite{chen2019chronoscope, petrelli2014photo}, music \cite{kim2019muredder, odom2020exploring, park2025nore}, and social media data \cite{duel2018supporting}. A parallel line of work has explored quantified self-tracking, where people reflect by examining historical records of sensor-based data \cite{elsden2016quantified, rooksby2014personal}. However, research in reflective informatics emphasizes that reflection does not arise automatically from exposure to personal artifacts or data; carefully scaffolded experiences are needed \cite{baumer2015reflective, slovak2017reflective}. 
To provide such scaffolding, early conversational agents explored structured reflection through rule-based interactions \cite{abdulrahman2023changing, klein2014intelligent, ly2017fully}. These systems have provided valuable entry points for reflection, though they often offered limited adaptability to individual contexts \cite{leusmann2024comparing}. 

More recently, Large Language Models (LLMs) have enabled personalized and adaptive support for reflection \cite{jo2023understanding, bae2022building}. ExploreSelf \cite{song2025exploreself} supports users in articulating personal challenges through adaptive questioning. In mental health contexts, MindfulDiary \cite{kim2024mindfuldiary} helps psychiatric patients document daily experiences through conversational journaling. For career exploration, Letters from Future Self \cite{jeon2025letters} facilitates introspection through dialogues with LLM-based agents that simulate a future self. These systems represent meaningful steps forward, yet they typically frame users as singular, unified selves. Given that the self can be viewed as dialogical and multi-positional, how to support exchanges among distinct internal perspectives remains underexplored. This motivates our approach to facilitate inner dialogue across multiple I-positions---a design goal we explore in this paper.

\subsection{Potential of Multi-Agent Systems for Supporting Inner Dialogues}
Recent studies have shown that LLMs are remarkably capable of simulating distinct personas \cite{cheong2022role, shao2023character}. Beyond surface-level role-playing, LLMs can now generate agents that reflect the complex interplay of identity components, such as personality traits \cite{jiang2023personallm, liu2024skepticism} and value systems \cite{zhou2023sotopia, xie2024can}, mirroring the richness of real-world individuals \cite{lee2025spectrum}. These agents could maintain consistent narratives and unique perspectives while responding adaptively to different contexts \cite{jeon2025letters}.

Building on this, multi-agent systems are emerging as powerful tools for surfacing diverse perspectives and domain-specific insights across fields \cite{guo2024large}. In particular, they have shown promise in collaborative problem-solving, decision support, and creative ideation \cite{su2024many}. For instance, in software engineering, agents simulating roles such as product managers, architects, and engineers can coordinate to tackle complex development tasks \cite{hong2024metagpt}. In creativity support, agents with distinct backgrounds and viewpoints can engage in brainstorming sessions, generating more original ideas than a single LLM operating alone \cite{lu2024llm}.

Here, we draw attention to the potential of combining LLMs' persona-simulation capabilities with a multi-agent architecture to support inner dialogue. By simulating distinct personas as different I-positions, such systems may capture complex dialogical processes---including disagreement, perspective shifts, and finding common ground---forms of interaction that prior studies have documented among agents within multi-agent structures \cite{xu2025autocbt, ryu2025cinema}.

Despite this potential, the use of LLM-based multi-agent systems for introspection through inner dialogue remains largely unexplored \cite{divekar2024externalizing, fang2025leveraging, fang2025mirai}. In response, we developed \textbf{InnerPond}, a research probe that applies a multi-agent LLM architecture to externalize inner voices as dialogical agents, enabling users to engage them in structured inner dialogue.

\section{InnerPond: Translating DST into Interactive Experience}

\begin{figure*}[!ht]
  \centering
  \includegraphics[width=0.85\textwidth]{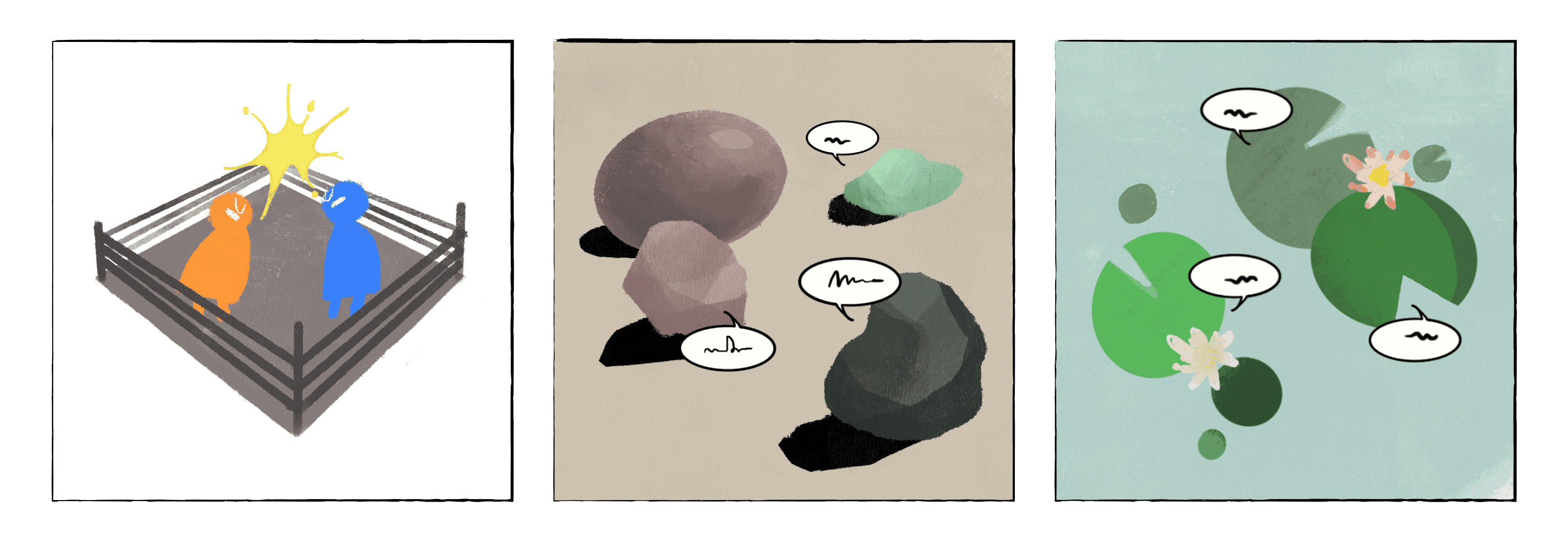}
  \caption{Three-phase metaphor evolution for the dialogical space: From group chat as battle ring (Phase 1), to stones highlighting individuality yet emphasizing separation (Phase 2), to lotus leaves appearing distinct yet sharing roots, enabling a meta-position perspective (Phase 3).}
  \Description{Three-phase metaphor evolution showing: a group chat room resembling a battle ring (left), stones arranged on a landscape representing individual I-positions (center), and a pond with lotus leaves sharing roots underwater (right).}
  \label{fig3}
\end{figure*}

Because Dialogical Self Theory (DST) conceptualizes the self as a plurality of interacting inner voices rather than a single unified identity, translating this theoretical perspective into an interactive system demanded careful attention to the experiential conditions under which inner dialogue can unfold. Our design process followed a research-through-design approach \cite{zimmerman2007research, zimmerman2014research}, using iterative prototypes as probes to explore how DST could be materially instantiated and how users might meaningfully engage with their plural selves.

At the outset, we established three core design goals grounded in DST's foundational principles. These goals functioned not as technical requirements but as orienting principles that guided design experimentation and decision-making throughout the development of InnerPond.

\textbf{\textit{Design Goal 1 (DG1) --- Support coexistence rather than resolution.}}
Inner conflict is commonly framed as a debate in which one position must ultimately prevail \cite{nir2012voicing}. In contrast, DST emphasizes coexistence, negotiation, and the ability to hold tensions without prematurely collapsing them into a final answer \cite{hermans2014self, hermans2011handbook}. Our goal was to create interaction structures that sustain multiple voices in parallel and resist convergence toward a single dominant perspective.

\textbf{\textit{Design Goal 2 (DG2) --- Ensure equal attention and legitimacy for each inner voice.}}
Plurality is not simply the presence of many voices; it requires that each voice be heard, recognized, and granted the opportunity to contribute meaningfully \cite{elliott1997multiple, raggatt2000mapping}. We sought to design a system that encourages engagement with both dominant and quieter perspectives, supporting self-authorship and preventing strong voices from overshadowing vulnerable or easily overlooked ones.

\textbf{\textit{Design Goal 3 (DG3) --- Foster relational connectedness and meta-positional perspective.}}
Externalizing multiple voices risks fragmenting the self into disconnected pieces. DST emphasizes that inner perspectives are dynamically related, and that reflection emerges through perceiving these relationships from a higher-level meta-position \cite{gkantona2023my, zhu2022positioning}. Our goal was to support users in experiencing a holistic, relational structure of identity---shifting from being inside a conflict to observing and orchestrating it from a meta-positional perspective.

These three goals served as the conceptual backbone of our design inquiry. The subsequent sections detail how iterative explorations of spatial metaphors, interaction scaffolding, and conversational orchestration functioned as design experiments to operationalize and balance these goals, culminating in the current form of InnerPond.

\subsection{Metaphor Evolution for the Dialogical Space}

Guided by these design goals, we explored spatial metaphors for a dialogical space where multiple inner voices could coexist as distinct yet interconnected agents. This exploration evolved through three phases---each iteration revealing tensions that prompted further refinement.

\textit{\textbf{Phase 1: Group Chat Room as Battle Ring.}} In the early design phase, we envisioned a `debate arena' based on the observation that inner conflict often manifests as competing perspectives \cite{nir2012voicing}. Borrowing the familiar group chat interface \cite{karahodvza2025conceptual, mogharrab2020family}, we envisioned creating an environment where each voice could clash in real-time. While this approach was promising for externalizing conflict, we realized that it implied the assumption that ``\textit{one voice must win}''---directly contradicting \textbf{DG1}'s emphasis on coexistence without premature resolution. This prompted us to seek a space where multiple perspectives could evolve together, rather than compete for dominance.

\textit{\textbf{Phase 2: Landscape of Mind through Stone Metaphor.}} To design a space for coexistence rather than victory , we drew inspiration from DST-based therapies  \cite{jahn2018using, van2010collage} that use natural objects like stones \cite{konopka2018composing} to symbolize I-positions and reconstruct identity through spatial arrangement. Translating this into a digital environment, we adopted a ``\textit{Landscape of Mind}'' interface where users could arrange digital stones to construct their own inner maps. This approach addressed \textbf{DG1} by allowing voices to coexist spatially, and partially supported \textbf{DG2} by giving each I-position a distinct, tangible representation. However, while the stone metaphor effectively highlighted the individuality of each I-position, we recognized that it simultaneously emphasized separation and fixed identity---failing to address \textbf{DG3}'s emphasis on relational connectedness and meta-position perspective \cite{hermans2010m, hermans2011handbook}.

\textit{\textbf{Phase 3: Inner Pond through Lotus Leaf Metaphor.}} To address the limitations of stones as representations of selves, we explored alternative metaphors, including a `Zen Garden' \cite{nitschke1993japanese, hermans2001dialogical}. However, its emphasis on stillness and fixed balance was misaligned with the dynamic, fluid nature of I-positions. We ultimately arrived at the ``\textit{Lotus Leaf}'' metaphor, where each leaf appears independent on the surface while remaining connected through shared roots beneath the water---capturing both the distinctness and interconnectedness of I-positions. This metaphor integrates all three design goals: leaves coexist on the pond surface without one dominating another (\textbf{DG1}); each leaf maintains a distinct and visible identity (\textbf{DG2}); and the shared underwater roots, together with the user's bird's-eye perspective, naturally embody DST's concept of the `\textit{meta-position},' enabling relational connectedness and holistic observation (\textbf{DG3}).

\subsection{Scaffolding for Inter-Self Introspection}
With the spatial metaphor in place, we shifted our focus to structuring how users would move through the introspective experience. As a reference framework, we drew on ``\textit{Composing the Self}'' \cite{konopka2018composing}, a DST-based therapeutic technique using stones described earlier. This therapy guides clients through a seven-step process---including 1) selecting stones as I-positions, 2) arranging them spatially to visualize relationships, 3) labeling (naming) each stone, 4) examining the arrangement, 5) voicing each position, 6) repositioning stones, and 7) reflecting on the composition as a whole.

We iteratively adapted these seven steps into four key stages, designed to follow a natural cognitive flow while lowering the barrier to engaging with one's multiple selves. Rather than preserving each therapeutic step as a discrete phase, we reorganized them into higher-level stages that reflect how these activities unfold in an interactive, AI-mediated context.

Specifically, Steps 1 (selecting stones) and 3 (labeling stones) were combined into \textit{Stage 1: I-position Construction}, while Steps 2 (arranging), 4 (examining), and 6 (repositioning stones) were consolidated into \textit{Stage 2: Spatial Configuration}. Step 5 (voicing each position) was adapted as \textit{Stage 3: Dialogical Exchange}, and Step 7 (reflecting on the composition) became \textit{Stage 4: Reflective Snapshot}. The following subsections elaborate on how each stage reinterprets its corresponding therapeutic steps by leveraging the affordances of an LLM-enabled, interactive medium.

\textit{\textbf{From Manual Selection to AI-Augmented Identification.}} The original therapy we drew on \cite{konopka2018composing} requires clients to structure and verbalize their inner world from scratch, demanding a high level of self-awareness and cognitive load \cite{jahn2018using}. We envisaged transforming this into AI-augmented identification, leveraging the analytical capabilities of LLMs. Specifically, we utilized LLMs to assist users in identifying their multiple selves, drawing on user-provided data such as personality traits, values, and personal narratives. To avoid imposing a fixed interpretation, we designed these AI-generated I-positions as starting points that users can enrich through co-creation---exploring, editing, or conversing with the suggested inner voices in their own words. This approach served as the foundation for \textbf{[Stage 1: I-position Construction]} in InnerPond, helping users concretize ambiguous inner voices into tangible objects (individual lotus leaves) and recognize their multifaceted selves.

\textit{\textbf{From Physical Arrangement to Digital Visualization.}} In the original therapy \cite{konopka2018composing}, clients are asked to express relationships between different selves through physical distance and height of stones. We translated this into digital manipulations, such as adjusting the position, size, and color of the lotus leaves. Unlike physical stones, these attributes can be easily adjusted, allowing users to fluidly explore and reshape their inner relationships. This approach was implemented as \textbf{[Stage 2: Relational Positioning]}, where users spatially compose their inner landscape by arranging lotus leaves on the pond.

\textit{\textbf{From Role-playing to Dialogical Exchange.}} In the original therapy \cite{konopka2018composing}, clients sequentially role-play each stone's perspective to negotiate inner conflicts. We reinterpreted this role-playing step as real-time dialogue among agents representing different selves as lotus leaves. This aimed to allow users to observe conversations among their selves from a third-person perspective or intervene as a mediator, without the burden of direct role-playing. This approach became the core of \textbf{[Stage 3: Dialogical Exchange]}, enabling dynamic exchanges among conflicting selves through dialogue.

\textit{\textbf{From Closing to Temporal Integration.}} In the original therapy \cite{konopka2018composing}, the session concludes with a one-time reflection on the final composition. We reframed this closing not as an ending but as a record of a ``\textit{Temporal Self-portrait}''---a record that can accumulate over time, helping users understand their inner world as a flowing narrative rather than a fixed entity. This concept was materialized in \textbf{[Stage 4: Reflective Snapshot]}, where users capture the current configuration of their pond to preserve the moment of introspection.

\subsection{Conversational Orchestration and System Implementation}

\begin{figure*}[!ht]
  \centering
  \includegraphics[width=0.9\textwidth]{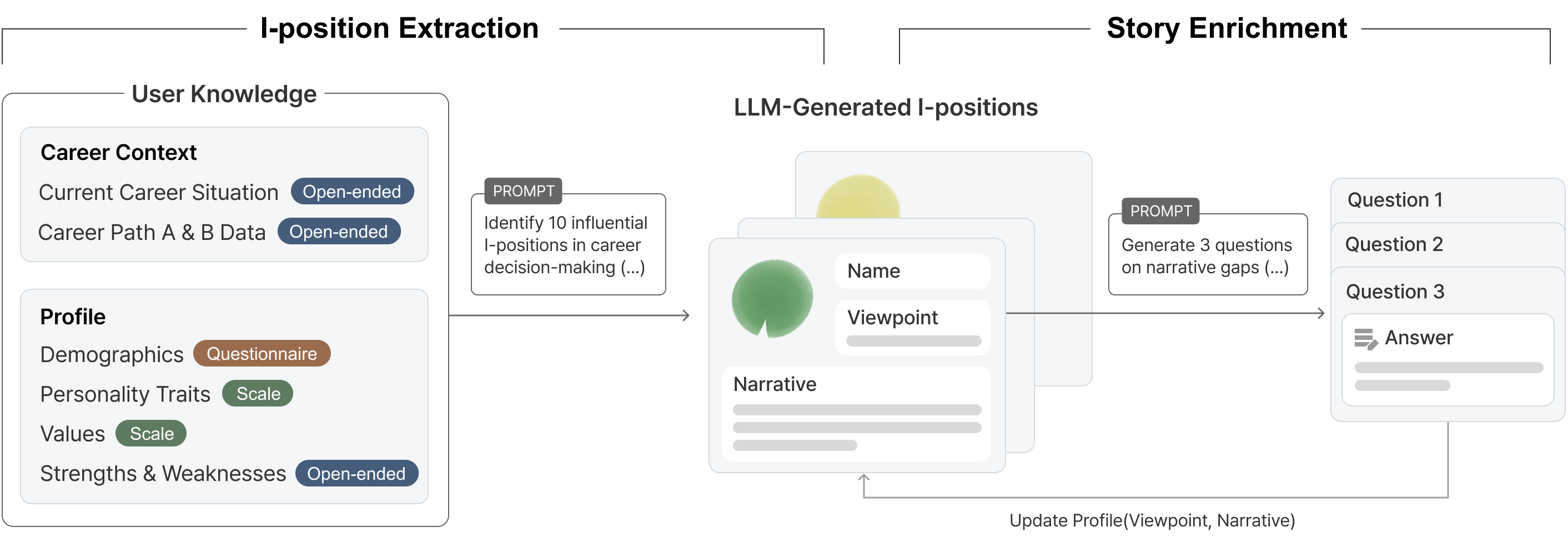}
  \caption{I-position extraction and story enrichment pipeline: user knowledge is transformed into distinct I-positions, iteratively refined through scaffolding questions and user responses.}
  \Description{This figure illustrates the I-position extraction and story enrichment pipeline. In the extraction stage, user knowledge (career context and profile data) is transformed into distinct I-positions, each composed of a name, viewpoint, and narrative. In the enrichment stage, the system generates scaffolding questions to identify narrative gaps. The user's responses then iteratively refine each I-position's viewpoint and narrative.}
  \label{fig4}
\end{figure*}

\begin{figure*}[!ht]
  \centering
  \includegraphics[width=0.75\textwidth]{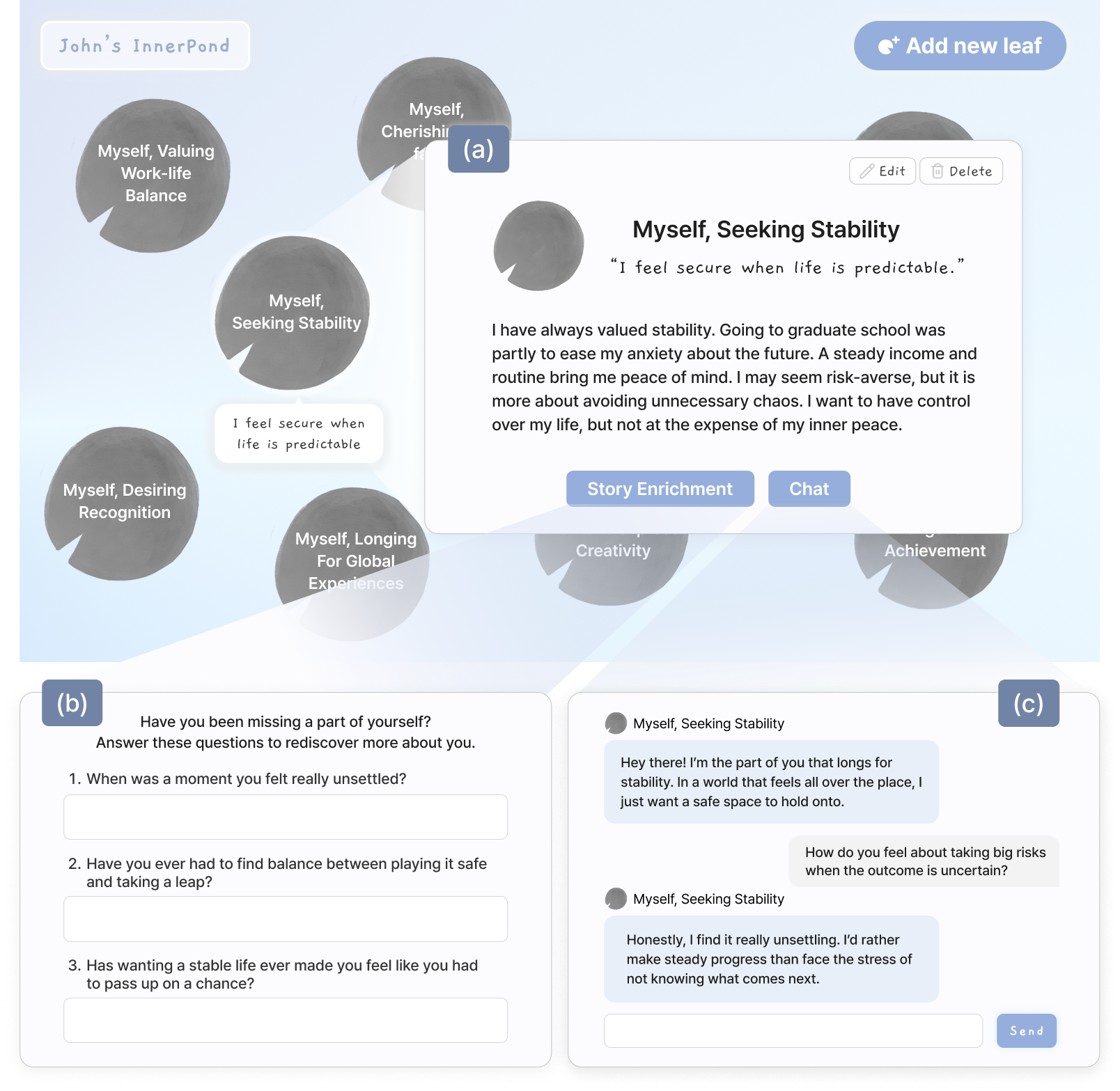}
  \caption{The interface for [Stage 1: I-position Construction]: (a) I-positions visualized as lotus leaves on the main pond view, with a profile modal showing name, viewpoint, and narrative; (b) ``Story Enrichment'' modal with scaffolding questions to refine the narrative; (c) ``1:1 Dialogue'' modal for direct conversation with a leaf agent.}
  \Description{This figure illustrates the Stage 1 (I-position Construction) interface. (a) The main pond view displays I-positions as lotus leaves; clicking a leaf opens a profile modal showing the I-position's name, core viewpoint, and narrative. (b) The ``Story Enrichment'' modal presents three scaffolding questions to help users refine the I-position's narrative. (c) The ``1:1 Dialogue'' modal shows a conversation between the user and a leaf agent, allowing users to explore the I-position's motivations.}
  \label{fig5}
\end{figure*}

The four-stage scaffolding established above was implemented in \textbf{InnerPond}, an LLM-based multi-agent system that guides users through four stages: \textit{(1) I-position Construction}, \textit{(2) Relational Positioning}, \textit{(3) Dialogical Exchange}, and \textit{(4) Reflective Snapshot}. While presented sequentially, these stages can be navigated freely, allowing for non-linear self-exploration.

We situated this implementation within the context of \textit{career decision-making}---specifically, situations where users face a choice between two career paths. Such decisions naturally evoke inner dialogue as multiple perspectives and value conflicts come into play \cite{ran2023linking}. Career decisions provide concrete scenarios in which diverse I-positions are activated, interact, and evolve. While our study focuses on this context, the system's architecture can extend to other situations involving inner multiplicity, such as work–life balance or relationship conflicts \cite{mitra2021life}.

In the following sections, we detail how each stage translates into concrete interface features and underlying technical mechanisms, and describe the overall technical configurations.

\begin{figure*}[!ht]
  \centering
  \includegraphics[width=0.65\textwidth]{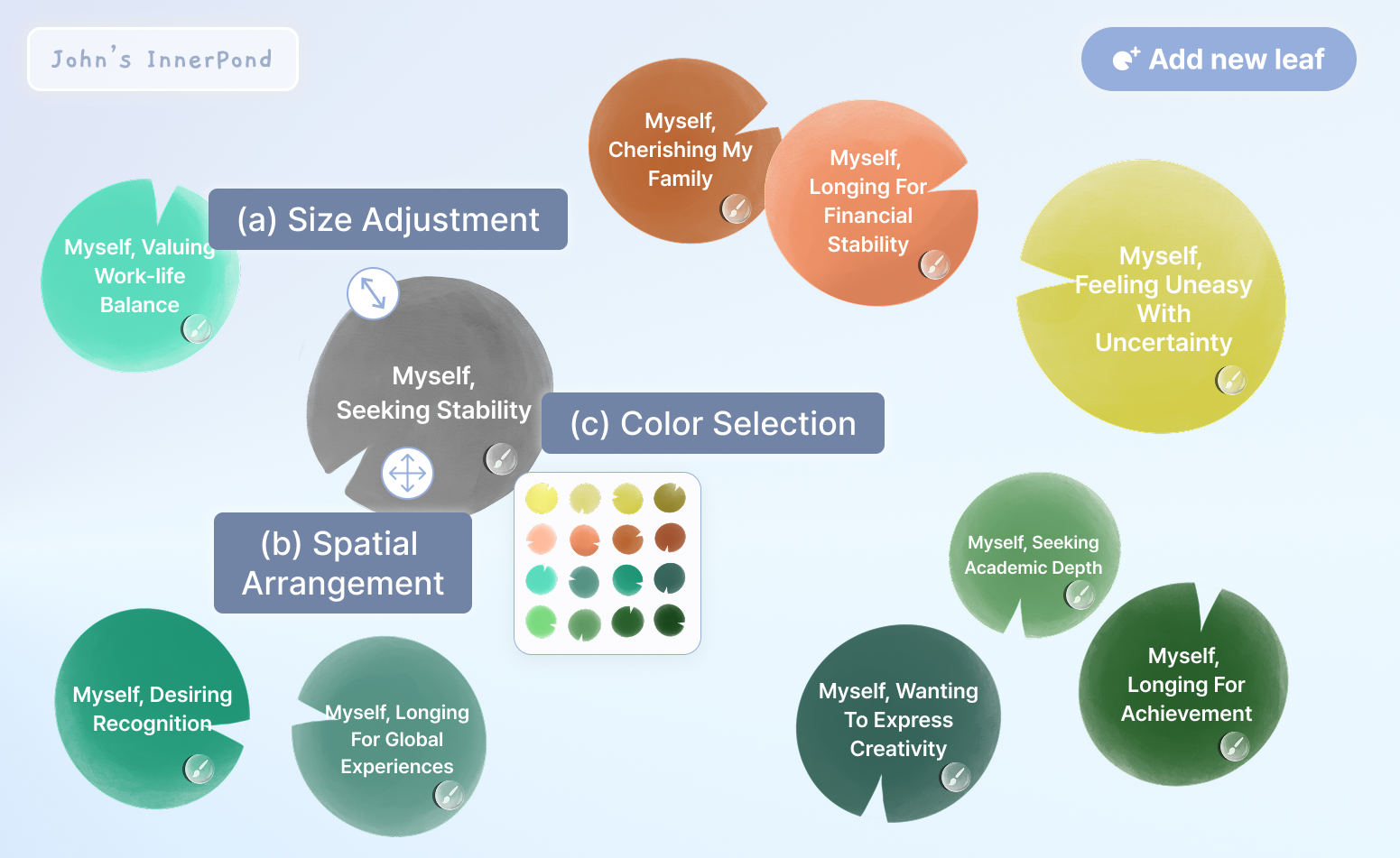}
  \caption{The interface for [Stage 2: Relational Positioning]: Users can freely customize their inner landscape by (a) adjusting leaf size, (b) repositioning leaves via drag-and-drop, and (c) selecting colors through the dewdrop-shaped button.}
  \Description{This figure illustrates the Stage 2 (Relational Positioning) interface. (a) Users can adjust a leaf's size by hovering over its edge. (b) Leaf positions can be modified via drag-and-drop to arrange the inner landscape. (c) Clicking the dewdrop-shaped button opens a color palette for personal expression. These manipulations allow users to visually compose their inner structure in personally meaningful ways.}
  \label{fig6}
\end{figure*}

\subsubsection{Stage 1: I-position Construction}
This stage centers on generating and enriching I-positions through a pipeline that combines LLM analysis with user input (Figure \ref{fig4}). To extract diverse selves relevant to users choosing between two career paths, we carefully designed a knowledge base structure using the SPeCtrum framework \cite{lee2025spectrum}. This knowledge includes demographics, personality traits (Big Five Inventory–2 Short Form (BFI-2-S) \cite{soto2017short}), work values (Super's Work Values Inventory (SWVI) \cite{robinson2008psychometric}), self-identified strengths and weaknesses, and the two career options under consideration. Quantitative scale data are converted into natural language summaries using Chain of Density (CoD) prompting \cite{serapio2023personality} for effective LLM understanding (see Appendix \ref{app:knowledge} for the complete knowledge structure).

Based on this knowledge structure, we prompted the LLM to extract approximately ten distinct inner voices (I-positions), ensuring no overlap among them. We ensured they were distributed evenly across three categories: (1) those common to both career paths, (2) those specific to one career path, and (3) those specific to the other. 
To represent each I-position as an authentic inner voice, we constructed a profile for each, comprising a name in the ``Myself, ...'' format (e.g., [Myself, Seeking Stability]), a core viewpoint capturing its unique voice, and a first-person narrative revealing underlying motivations (see Appendix \ref{app:initial} for the full prompt).

Based on these profiles, each I-position is visualized as an interactive leaf agent---a named lotus leaf on the pond that embodies a distinct inner voice (Figure \ref{fig5}-(a)). Users can hover over a leaf to reveal its core viewpoint or click to open a modal with the full profile. To encourage equal attention to all voices, all leaves initially appear in the same gray color and size.

To prevent users from passively accepting AI-generated I-positions, we introduced a co-creation process that allows users to refine and personalize them. The ``\textit{Story Enrichment}'' feature generates three scaffolding questions by identifying gaps in the current narrative (see Appendix \ref{app:enrich-q} for the full prompt). As users answer these questions, the leaf agents are enriched with the user's own voice, reflecting their specific context rather than a generic description (Figure \ref{fig5}-(b)). Users can also modify I-positions through direct editing or add and delete leaf features.

Finally, users can engage in ``\textit{1:1 Dialogue}'' with each leaf agent (Figure \ref{fig5}-(c)). This dialogue allows users to question, challenge, or explore an I-position's motivations without interference from other voices. Each leaf agent maintains its core identity to express consistent perspectives even when challenged (see Appendix \ref{app:one-on-one} for the full prompt).

\subsubsection{Stage 2: Relational Positioning}
This stage focuses on externalizing the relationships among I-positions, allowing users to spatially compose their inner landscape. Once users have explored and enriched their I-positions, they could begin to arrange them spatially to express relationships among their inner voices (Figure \ref{fig6}). Users can freely adjust the position, size, and color of each leaf. Unlike physical stones, these attributes can be easily changed, allowing users to fluidly explore and reshape their inner landscape in personally meaningful ways.

\begin{figure*}[!ht]
\centering\includegraphics[width=0.9\textwidth]{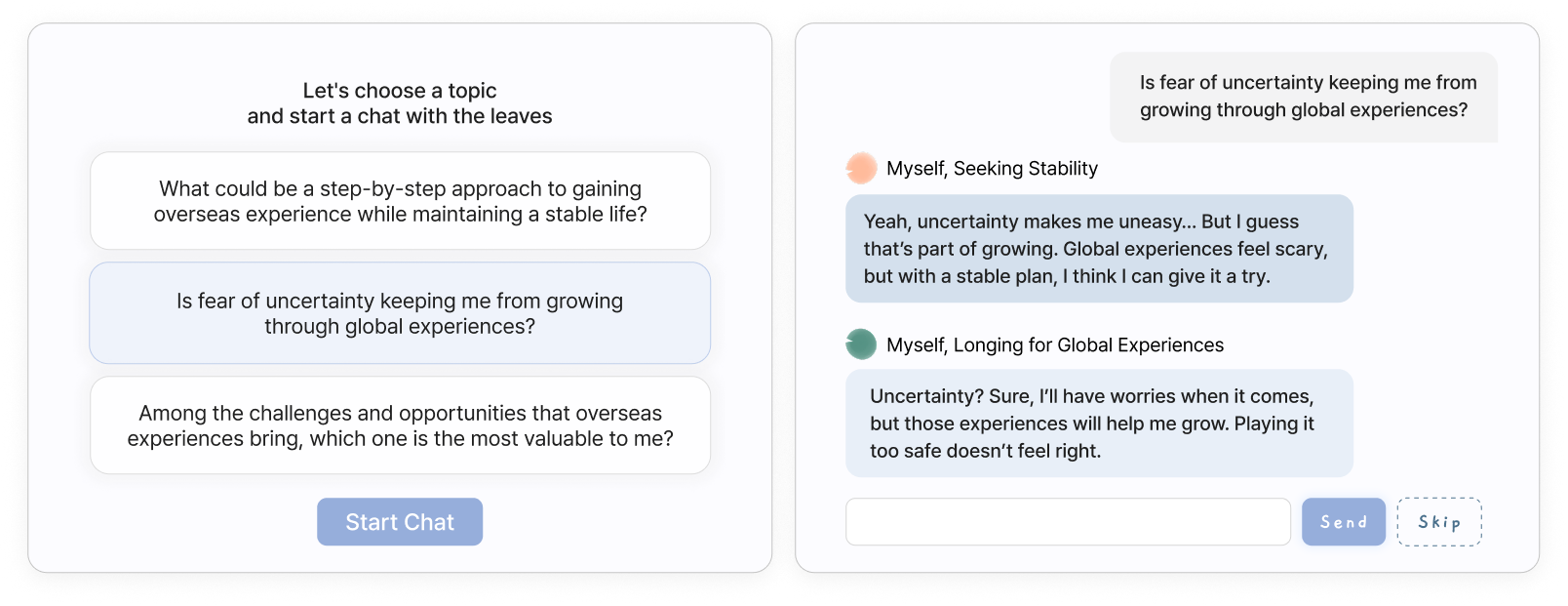}
  \caption{The interface for [Stage 3: Dialogical Exchange]: (Left) After selecting two leaf agents, the system suggests three discussion topics based on their relationship; (Right) the group chat where users can participate via ``Send'' or observe via ``Skip.''}
  \Description{This figure illustrates the Stage 3 (Dialogical Exchange) interface. (Left) After selecting two leaf agents, the user chooses one of three tailored discussion topics to initiate the conversation. (Right) The group chat shows two leaf agents responding to the selected topic from their distinct perspectives; the user can participate by sending a message or skip to observe the dialogue between agents.}
  \label{fig7}
\end{figure*}

\subsubsection{Stage 3: Dialogical Exchange}
This stage aims to facilitate dynamic dialogue among I-positions, allowing users to observe or mediate conversations between their inner voices. This multi-agent conversation begins when users select two leaf agents for dialogue (Figure \ref{fig7}-Left). The LLM then analyzes their relationship and generates three tailored discussion topics based on their dynamics (see Appendix \ref{app:mti-topic} for the full prompt):

\begin{itemize}
\item \textit{Conflict}: Questions to explore value clashes, seek compromise and navigate dilemmas.
\item \textit{Complementary}: Questions on how distinct aspects can work together.
\item \textit{Unrelated}: Questions to explore the diversity and unique motivations of each self.
\end{itemize}

After a user selects a topic, the group conversation begins and is coordinated by an orchestrator module implemented through prompt engineering rather than a separate multi-agent framework. The orchestrator functions as a control prompt that tracks conversational context and interaction state, managing turn-taking among the two leaf agents and the user to prevent any single agent from dominating. For example, when a direct question is posed, the addressed agent is prompted to respond first; when a challenging argument arises, the orchestrator issues a targeted prompt inviting the opposing agent to refute.

Users can engage in the conversation through two modes (Figure~\ref{fig7}-Right). In observation mode, activated via the \textit{Skip} feature, the system injects an intervention message—``Do not repeat viewpoints; engage more deeply with each other’s perspectives''—to encourage sustained dialogue between agents. In mediation mode, users intervene directly through their own input; the agents then respond while maintaining their core identities, allowing users to guide the direction of the dialogue.

\subsubsection{Stage 4: Reflective Snapshot}
This stage intends to allow users to capture a snapshot of their current inner landscape. By clicking the \textit{save pond} button, users can capture their current pond configuration---including leaf arrangement and visual attributes---as a timestamped image file (e.g., ``\{user\}'s InnerPond\_\{timestamp\}''). Users can revisit past inner landscapes at any time, helping them view the self not as fixed but as an evolving narrative.

\subsubsection{Technical Configuration}
InnerPond was implemented as a web-based application using TypeScript \cite{microsoft2025typescript} and Next.js framework \cite{vercel2025nextjs}. The system consisted of a frontend for user interactions and a backend managing user data and LLM APIs. All interaction data was stored in MongoDB \cite{mongodb2009}, with access restricted to the research team only. For the generative pipeline, we utilized Anthropic's Claude-3.5-Sonnet, selected for its strong performance in Korean among models available at the time of the study \cite{kim2024click, jang2024evaluating}.

The system's generative functionality is organized into three LLM-driven pipelines, each implemented through prompt engineering rather than a dedicated multi-agent framework. First, \textit{I-position Extraction} generates initial, personalized I-positions from pre-survey data. Second, \textit{Single-Agent Interactions} support story enrichment and one-to-one dialogue between the user and an individual leaf agent. Third, \textit{Multi-Agent Orchestration} coordinates dialogue among multiple leaf agents by using a control prompt that manages turn-taking and interaction flow based on conversational context. 
Selected prompts are included in the Appendix \ref{app:pipeline1}-\ref{app:pipeline3}; full prompts are available at \url{https://github.com/syou-b/innerpond}.

\section{Methods}

To explore how people experience and make sense of engaging with their multiple selves (RQ2), we conducted a qualitative exploration with participants actively deliberating between two career paths, using InnerPond as a research probe for enabling inter-self dialogue.

\subsection{Participants}
We recruited 17 participants (7 male, 10 female, average age = 27.18 (\textit{SD}=4.93, Min=21, Max=37)) who were actively deliberating between two career directions, providing a naturally high-stakes context rich in internal conflict and self-negotiation. Recruited via university communities, participants were either (a) undergraduate or graduate students expected to graduate within three years, or (b) recent graduates currently exploring career opportunities while unaffiliated with any educational institution or workplace.

\subsection{Study Procedure}
The three-week study involved three sessions: (1) pre-survey, (2) in-person session, and (3) follow-up interview. All research materials and procedures were approved by the Institutional Review Board (IRB) of the hosting university. The entire study, from materials and sessions to data analysis, was conducted in Korean. For the analysis and writing process, participants' quotes were translated into English by bilingual researchers on our research team.

\subsubsection{Pre-survey}
To construct knowledge of participants' multiple selves relevant to their career deliberations, participants completed an online pre-survey seven days prior to the in-person session. The survey consisted of two main parts:

\begin{itemize}
\item \textbf{Personal Information Survey (5 min)}: Participants provided demographics and completed personality assessments using the BFI-2-S \cite{soto2017short} and work values assessment through the SWVI \cite{robinson2008psychometric}. Participants were also asked to list three personal strengths and weaknesses.
\item \textbf{Career Context Survey (5 min)}: Participants briefly described the two career paths under consideration through short, open-ended responses. For each path, they detailed: (1) the background and rationale for considering it, (2) their expectations and concerns, and (3) any preparation processes undertaken.
\end{itemize}

This information was used to generate agents representing participants' I-positions prior to the in-person session (see Appendix \ref{app:knowledge} for an example).

\subsubsection{In-person Session}
We conducted the in-person session in a quiet room with two researchers for approximately 120 minutes. To foster a relaxed and introspective atmosphere, the lab space was set with meditative music and dim lighting. During the session, one researcher guided participants and conducted interviews (facilitator), while the other observed participant behavior and system interactions in real-time (note-taker), documenting notable moments and emerging patterns in field notes.

As the session began, participants provided informed consent after being briefed on the study's objectives and procedures. To allow the observing researcher to monitor interactions without disrupting the participant, participants shared their screens via Zoom. 
The seven-stage session---comprising an initial interview, onboarding, four InnerPond activities each followed by a post-activity interview, and an exit interview---was designed to investigate participants' evolving experiences. 
In particular, post-activity interviews explored participants' in-the-moment experiences by probing specific decisions and interaction patterns observed during each activity. To facilitate this, the facilitator used the documented field notes as prompts to help participants recall their process, jointly revisiting moments of interest to understand the reasoning behind their actions.
All interviews were conducted using a semi-structured interview guide, with probing and follow-up questions informed by participants' responses and observed interaction patterns (see Appendix \ref{app:interview}).

\begin{enumerate}
    \item[\textbf{(1)}] {\textbf{Initial Interview (5 min)}}: We began with a brief interview to understand participants' current career concerns and how they typically engaged in inner dialogue. 
    \item[\textbf{(2)}] {\textbf{Onboarding (5 min)}}: We introduced InnerPond, explaining that their inner voices---extracted from the pre-survey---were represented as `lotus leaves.' We presented an overview of the four activity stages, along with guidelines for the main features of each stage. Participants were informed that interviews would follow each activity and that researchers would be available to assist them.
    \item[\textbf{(3)}] {\textbf{I-position Construction (25 min activity + 15 min interview)}}: After onboarding, participants were asked to examine the extracted leaves in their InnerPond and articulate them. The post-activity interview explored the authenticity of the extracted I-positions and their motivations for editing, adding, or removing leaves (e.g., \textit{``Which leaf did you feel represented you the best, and why?''}, \textit{``Why made you delete the [Myself, ...] leaf?''}).
    \item[\textbf{(4)}] {\textbf{Relational Positioning (5 min activity + 10 min interview)}}: Participants were instructed to visually express their inner landscape by adjusting the size, color, and position of leaves in the pond, while thinking-aloud to verbalize their thoughts. The post-activity interview explored the rationale behind their pond structure and design, probing specific choices observed during the activity (e.g., \textit{``When arranging the leaves, what did you consider and why did you choose their positions?''}, \textit{``What made you not resize the lotus leaves at all?''}). 
    \item[\textbf{(5)}] {\textbf{Dialogical Exchange (15 min activity + 10 min interview)}}: Participants were instructed to create and engage in group conversations with different leaf combinations and topics. The post-activity interview focused on their group conversation experience, their criteria for selecting leaves, and the suitability of LLM-suggested topics, as well as observed conversation patterns (e.g., \textit{``Why did you focus most of your time on conversations in this group chat?''}).
    \item[\textbf{(6)}] {\textbf{Reflective Snapshot (10 min activity + 5 min interview)}}: In the final activity, participants freely explored previous stages at their own pace, revisiting and engaging with their I-positions as they wished. Afterward, participants were asked to save their InnerPond landscape using the `save pond' feature. The brief post-activity interview explored what participants chose to revisit (e.g., \textit{``What did you mainly do during the free exploration time?''} and their experience of capturing a snapshot \textit{``How did you feel when you saved your pond at the end?''}).
    \item[\textbf{(7)}] {\textbf{Exit Interview (10 min)}}: After completing all stages, an exit interview was conducted about the overall InnerPond experience. We inquired about participants' most memorable moments, the perceived naturalness of the dialogue, and any new insights they gained. We also explored how the InnerPond experience impacted their self-understanding. Upon completion of the session, participants were compensated with 40,000 Korean Won (equivalent to 28 USD).
\end{enumerate}

\subsubsection{Follow-up Interview}
Two weeks after the in-person session, we conducted a follow-up phone interview to explore how participants reflected on their experiences with InnerPond following the session. Although participation in the follow-up interview was not mandatory, all but one of the 17 participants chose to take part. Following the semi-structured approach, the interview focused on whether the insights from the session persisted or evolved, whether participants noticed new patterns in how they engaged with their inner voices in daily life, and any shifts in their approach to career deliberation.

\subsection{Data Analysis}

\subsubsection{Data Preparation}
We documented participant behavior and experiences through multi-faceted data: (1) system logs and (2) qualitative data from in-person sessions and follow-up interviews. System logs provided detailed behavioral traces across all stages, including I-position profiles, their modifications, and chat messages with individual leaves (Stage 1), spatial configurations of the pond (Stage 2), and group conversation logs with leaf combinations and discussion topics (Stage 3) (see Appendix \ref{app:system_log} for detailed data description). Qualitative data included field notes documenting participant behavior and screenshots during activities, think-aloud transcripts from Stage 2, and all interview transcripts. All interviews were transcribed by the two researchers who conducted the study. We synthesized all data into a master document for each participant, integrating system logs and qualitative data. Two researchers collaboratively constructed these documents and visually organized the materials on FigJam \cite{staiano2022designing} to support systematic reference throughout the analysis.

\subsubsection{Thematic Analysis}
We employed thematic analysis \cite{braun2012thematic} to systematically analyze the collected data. Our approach combined deductive and inductive elements: we used the four stages of InnerPond as an overarching deductive framework to segment participants' experiences and behaviors. Within each stage, we performed open coding and inductively developed themes from the data. The analysis primarily operated at the semantic level, focusing on participants' explicit accounts of their experiences, while also attending to latent meanings when interpreting the underlying motivations and significance that participants attributed to their interactions. The entire research team conducted this analysis through weekly meetings, continuously cross-referencing insightful excerpts from interview transcripts with corresponding system logs to develop richer interpretations. Through an iterative process, we grouped the codes into higher-level themes and further developed them into more specific sub-themes to capture emergent patterns.

As a result, we identified key themes across the four stages: the externalization and personalization of inner voices (Stage 1), relational positioning through spatial composition (Stage 2), the orchestration of dialogue among multiple voices (Stage 3), and the preservation of temporal self-portraits (Stage 4). Additionally, through follow-up interviews, we observed an enduring theme of reflective integration into everyday life and career decisions.

\section{Findings}

This section reports how InnerPond, used as a research probe, shaped the ways participants engaged with their plural selves as they moved through its four designed stages. Sections 5.1--5.4 present findings from the in-person session, organized by each stage, while Section 5.5 reports insights from follow-up interviews conducted two weeks later. Rather than focusing on the stages themselves, we highlight the kinds of experiences that emerged within them---how participants articulated inner voices (Stage 1), made their relationships visible (Stage 2), encountered dialogue among these voices (Stage 3), and captured temporal self-portraits (Stage 4). Lastly, we describe how these experiences continued beyond the sessions and informed participants' everyday reflections and career decisions.

\subsection{Co-Creating I-positions Through Externalization and Reconstruction}
In Stage 1, participants co-created personalized I-positions from AI-generated fragments of self. Overall, externalizing their inner world into discrete `lotus leaves' allowed them to step back and view their psychological states from a distance, creating space for a deeper recognition of their multi-layered identity and the surfacing of overlooked aspects of themselves.  

\subsubsection{Initial Perceptions of Externalized Selves}
Participants' first encounters with the AI-generated leaves began with an immediate scan of how well each one aligned with their lived experience---whether it felt resonant, subtly off, or misaligned. Overall, most participants accepted the majority of their leaves, with over half (9/17) retaining the entire set, while others removed only one or two (\textit{M}=1.17). Many found the externalization itself valuable for introspection. P1 described how seeing her inner world visualized prompted an objective re-evaluation, while P8 found that having her unarticulated thoughts externalized allowed her to explore new self-perceptions, prompting reflections like, \textit{``Oh, I can think this way.''}

Many also described a sense of recognition as they scanned the leaves, noting how familiar worries or aspirations were suddenly given form. P7, for instance, was surprised to see her anxieties concretely represented: \textit{``Except for deleting one leaf, every single thing I always worry about was specifically targeted and realized. It felt like [they are] my actual inner voices.''} In particular, when the leaves reflected specific, personal experiences rather than generic concerns, participants received them with a pronounced sense of authenticity. For example, P11's [Myself, Rewarded by Admin Work] resonated deeply because the narrative reflected her conviction that \textit{``supporting the organization from behind is more meaningful.''} 

Others noted subtle misinterpretations rather than outright inaccuracy. P5 felt [Myself, Longing for the Prestige of Professorship] overstated her motivations: \textit{``I just thought of it as an honorable job. But this makes it sound like I'm thirsting for social recognition, and that's not quite right.''} Her reaction illustrated how even minor discrepancies in framing could shift participants' sense of ownership over a given I-position.

Overly misaligned leaves sometimes prompted resistance, causing some participants to remove them immediately. P7, for example, was decisive in removing [Myself, Anxious about Success]: \textit{``This is not me.''} Yet more often, participants chose to retain discordant leaves as prompts for reflection. P3 was initially skeptical of [Myself, Drained by Socializing] but later reframed her stance and chose to keep it: \textit{``At first, I didn't recognize it, but seeing it listed with the same weight [as others leaves] made me consider it more.''} Rather than dismissing the unfamiliar voice, she found that its presence surfaced neglected aspects of her career thinking---demonstrating how even imperfect extractions could also serve as catalysts for deeper self-examination.

\subsubsection{Elaborating and Personalizing the Extracted Selves}
After moving past their initial impressions, participants entered a more active phase of working with the leaves---clarifying, enriching, and reshaping the AI-generated narratives so they better reflected the nuances of their own experiences. In doing so, all participants engaged in story enrichment by answering AI-scaffolded questions (\textit{M}=11.82 times), to refine abstract or generic voices into situated, meaningful selves. Through this process, participants personalized the abstract leaves by grounding them in their own contexts---reshaping initially unconvincing or generic narratives to better align with their values and perspectives. For example, P3 developed the vague narrative of [Myself, Wanting to Contribute] into a clear professional aspiration. Answering tailored questions like ``What specific social problems do you want to solve?'' helped her move from an abstract desire to a concrete consideration of which issues she might want to address as a lawyer.

Participants also took direct control by modifying narratives that felt insufficient or misaligned. P6, for instance, edited the LLM's phrasing to better reflect her perspective, replacing \textit{``economic stability''} with \textit{``economic abundance.''} Furthermore, when participants felt important aspects of their inner voice were missing, they added new leaves to fill these gaps. In total, 12 participants added an average of 1.67 new leaves. P2, for instance, added [Myself, Valuing Office Environment], explaining how his happiness and efficiency depended on tangible workplace factors, like a comfortable atmosphere and good food. P4 went further, intentionally creating [Myself, Believing in Myself] to counterbalance the negative leaf [Myself, Anxious about Being Ordinary], as a move to prevent his inner landscape from being dominated by negative representations.

Through these processes of enrichment, modification, and addition, participants transformed the AI-generated leaves into personalized representations. The resulting leaves (\textit{M}=11.06) were not perceived as static data summaries but as distinct entities, each embodying a different aspect that constituted the self: \textit{``beings similar to me, but with a distinct character---like in the movie `Inside Out'''} (P15). In that film, emotions such as Joy and Sadness are portrayed as independent characters living inside a person's mind. This comparison suggests that the co-creation process gave the I-positions a sense of independent identity, making them feel like genuine inner voices rather than mere labels. Participants also valued the process for surfacing aspects of themselves they typically overlooked---even when the output was not perfectly accurate. P3 reflected: \textit{``With my single consciousness, I can't pay attention to everything at once. This process helped me extract parts of myself I usually miss. So even if some leaves felt off, it was still meaningful.''} These dynamics illustrate how co-creation shifted the system from simply classifying users to enabling a shared ownership of their plural selves.

\subsubsection{Conversing with Individual Selves}

\begin{figure}[t]
  \centering
  \includegraphics[width=0.9\columnwidth]{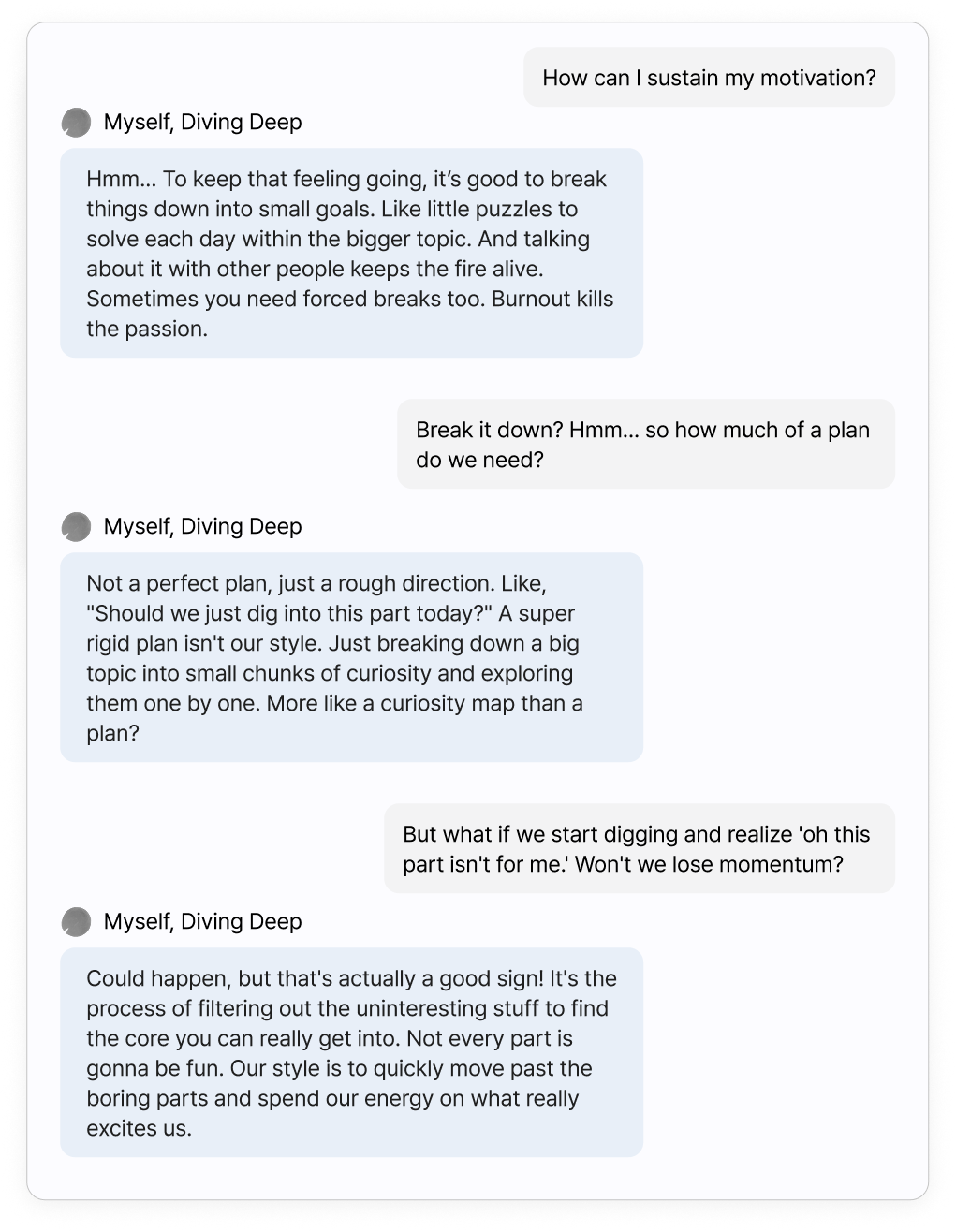}
  \caption{Example of a one-on-one dialogue with [Myself, Diving Deep] (P15).}
  \Description{This figure provides an example of a one-on-one dialogue between a user (P15) and their I-position, `Myself, Diving Deep.' In the conversation, the user asks for advice on sustaining motivation. The `Myself, Diving Deep' persona responds with insights that reflect its core character, suggesting a flexible, curiosity-driven approach over a rigid plan. This exchange demonstrates how users can engage with specific facets of their personality as distinct conversational partners to explore personal challenges and gain new perspectives.}
  \label{fig8}
\end{figure}

\begin{figure*}[!ht]
  \centering
  \begin{subfigure}{0.48\textwidth}
    \centering
    \includegraphics[width=\linewidth]{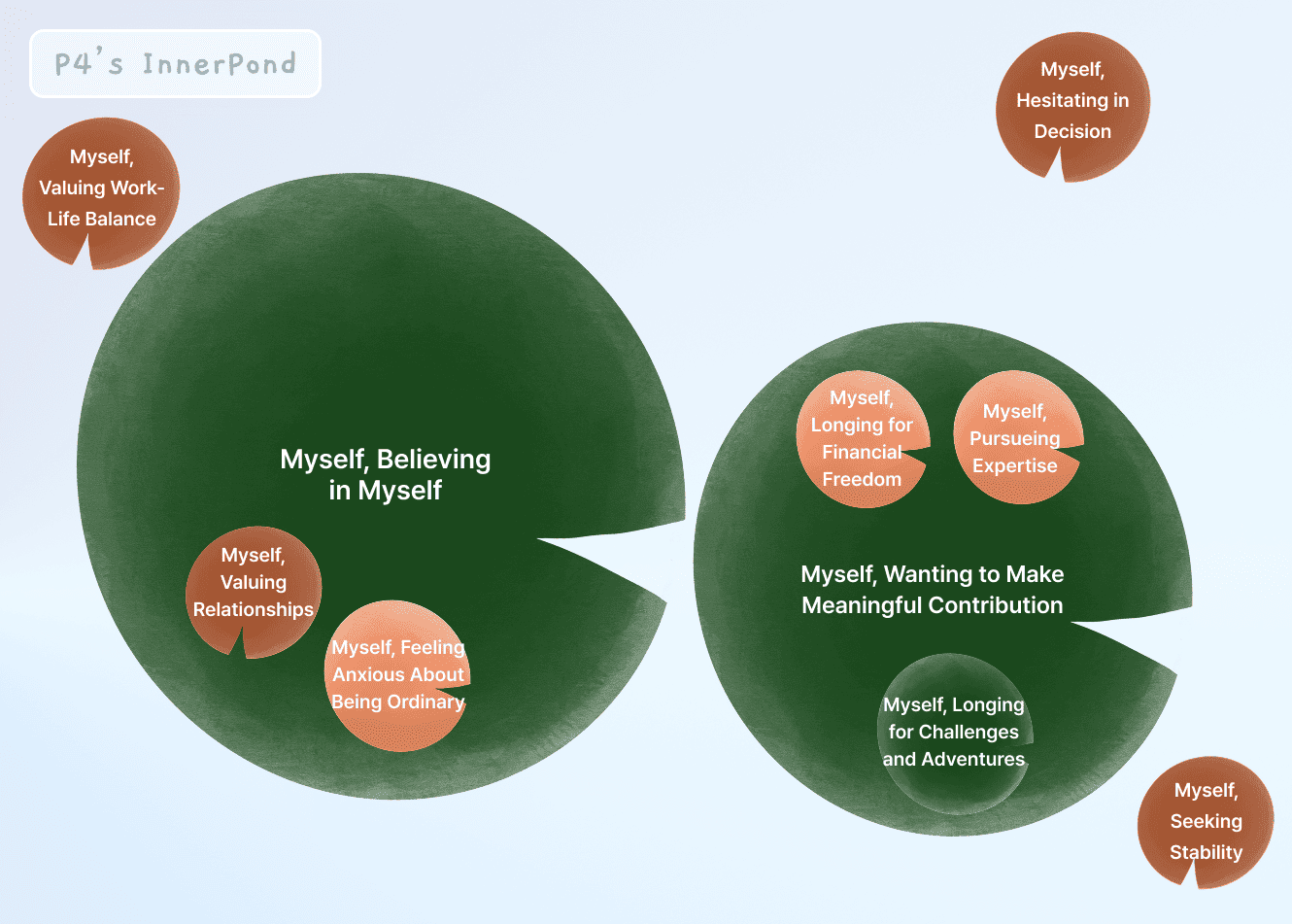}
    \caption{Inner landscape of P4}
    \label{fig9a}
  \end{subfigure}
  \begin{subfigure}{0.48\textwidth}
    \centering
    \includegraphics[width=\linewidth]{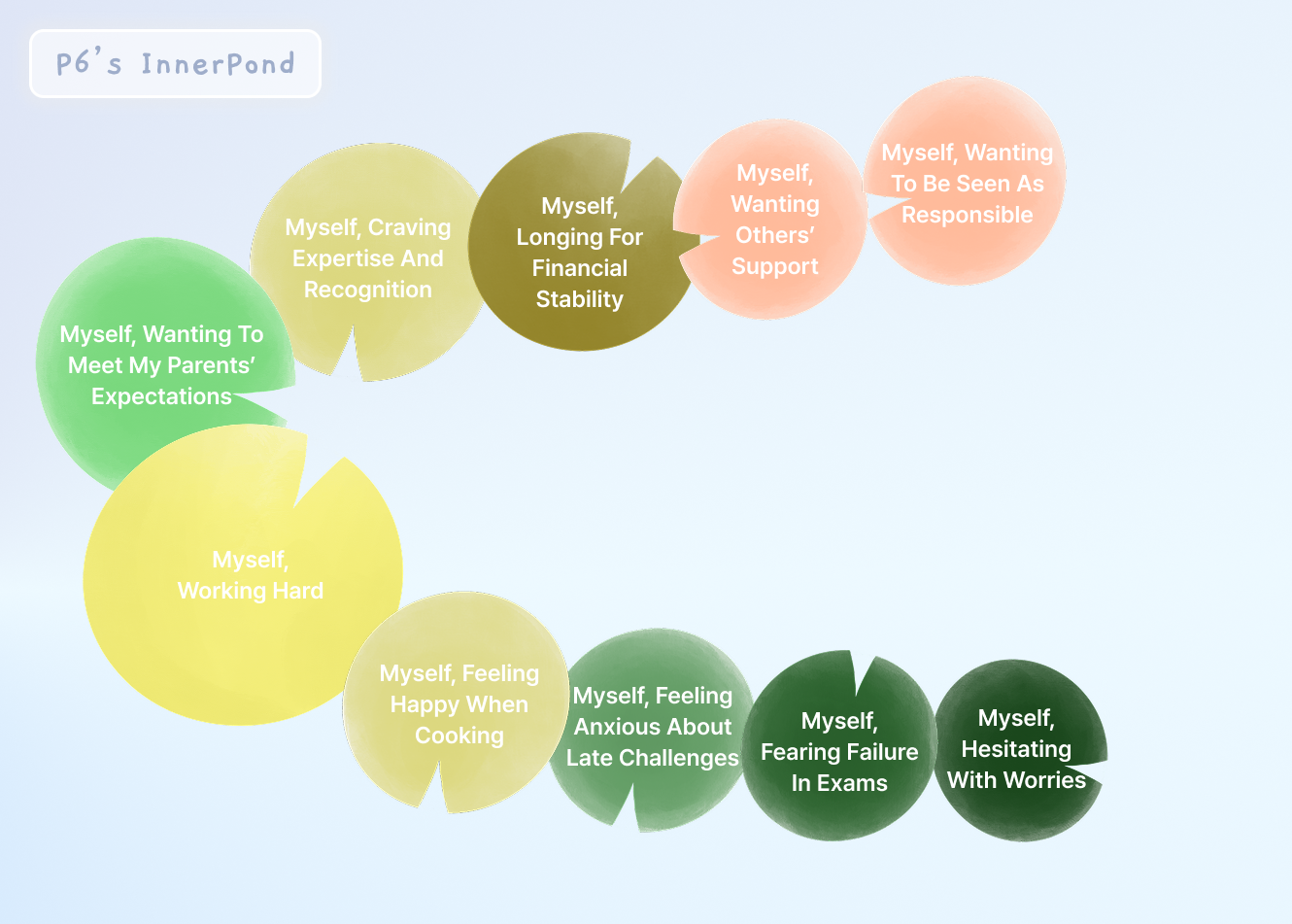}
    \caption{Inner landscape of P6}
    \label{fig9b}
  \end{subfigure}

  \caption{Examples of holistic inner landscapes composed by participants. (a) P4 created a scene of self-compassion, describing one self whispering encouragement to another. (b) P6 arranged the leaves to express gradual personal growth, likened to a musical crescendo.}
  \Description{This figure presents two examples of inner landscapes, demonstrating how participants used the size and spatial arrangement of I-position leaves to convey personal narratives. Participant P4 (a) depicted a scene of self-compassion by positioning a large `Myself, Believing in Myself' leaf behind `Myself, Wanting to Make Meaningful Contribution,' creating a visual impression of one self whispering encouragement to another. Participant P6 (b) arranged numerous leaves in a progressive spread of varying sizes to represent gradual personal growth, likening the composition to a musical crescendo.}
  \label{fig9}
\end{figure*}

With each I-position (leaf) elaborated and structured, the process shifted from organization to directly engaging these selves in one-on-one dialogue. Rather than conversing with all leaves, participants gravitated toward about half of their selves (\textit{M}=5.29), focusing on those where they felt tension, curiosity, or a need for clarity. These conversations took varied forms, shaped by participants' intentions and their relationship with each self.

Some participants used the one-on-one space for open-ended exploration, treating the self as a partner for thinking through complex issues. P15, for instance, held a collaborative conversation with [Myself, Diving Deep] to shape his research direction, asking questions like, \textit{``How can I find an interesting topic?''} and \textit{``How can I sustain my motivation?''} (Figure~\ref{fig8}) to facilitate open-ended reflection on the meaning of his research. Rather than seeking definitive answers, he used the dialogue to think aloud with a part of himself he wanted to understand better.

Others approached conversations more strategically. P7, for example, posed the same fundamental questions---like \textit{``Should I choose what I love or what I’m good at?''}---to a cluster of related selves, including [Myself, Fearing Uncertainty], [Myself, Pursuing Success], and [Myself, Wanting Recognition]. By collecting responses from multiple I-positions, she compared perspectives and analyze her career concerns from different angles.

The course of dialogue also depended on how familiar participants were with each self's voice---and how willing they were to acknowledge it. With highly aligned selves, conversations were often used to verify authenticity, sometimes even through adversarial testing. P13, for example, intentionally challenged [Myself, Craving Quick Success] by asking whether he should simply lower his expectations. When the self countered that doing so would only lead to future regret, he found he couldn't refute it: \textit{``I tried being uncooperative in the conversation, but the voice was perfectly aligned with what's inside me.''} By contrast, conversations with less familiar or initially rejected selves often unfolded as a process of gradual acceptance. P17 initially denied [Myself, Lacking Expertise], which voiced his professional insecurities. Yet through continued dialogue, he came to recognize a suppressed anxiety: \textit{``At first I kept denying it. But as we talked, I thought it could be a part of me. Maybe there has been a voice of anxiety in my heart, but I didn't know it because I didn't have the time or energy to think about it.''} 

As dialogues deepened, many participants also varied their conversational style to match the persona of each leaf. P7 discovered she was acting as a mediator, emphasizing possibilities when talking to selves expressing uncertainty, and highlighting risks when engaging with overly optimistic selves. For example, she encouraged [Myself, Anxious about Success] by reminding it that abilities can be developed, while cautioning [Myself, Passionate for Challenges] that excitement alone might not sustain long-term commitment. 

This adaptive approach was particularly effective in conversations with inherently negative selves, as it allowed participants to objectively view their weaknesses. P5 found it particularly \textit{``fun and helpful to talk with the leaves that contained my flaws,''} such as [Myself, Lacking Perseverance]. She explained that hearing her weaknesses spoken by another entity created emotional distance, making them easier to examine without self-judgment. P12 echoed this sentiment: through conversations, he was compelled to confront contradictions he would normally dismiss, describing it as \textit{``a powerful moment of reflection that came from seeing [his] weaknesses objectively.''} 

However, a few participants reported feeling fatigued when a negative self became too entrenched in its character. P11 felt drained after a conversation with [Myself, Tense in Relationships], noting that \textit{``if a self is too negative, it's tiring, even though it's me.''} P6 also described [Myself, Hesitant to Decide], as caught in an endless loop of anxiety: \textit{``Because it kept circling around anxiety and hesitation, it didn’t feel like it would offer a solution.''} These experiences suggest that while externalizing negative aspects could facilitate constructive self-understanding, overly repetitive portrayals may become counterproductive.

While most participants actively engaged in one-on-one dialogues and found them interesting and valuable, a few who were not accustomed to introspection found it difficult to initiate conversations with their selves. P4 explained: \textit{``I'm not the type to have thought deeply about myself. When I get a question, I can think about it, but when I tried to start a conversation, the questions didn't easily come to me.''} His account suggests that open-ended dialogical engagement presupposed a level of self-directed initiative that not all participants felt equipped to sustain.

\subsection{Composing Relational Landscapes Through Visual Arrangement}

In Stage 2, participants visually composed their `inner landscape' by arranging individual selves as lotus leaves. Building on the clearer understanding of each leaf developed in Stage 1, they moved beyond considering leaves in isolation and began attending to how different aspects of themselves related, contrasted, or clustered together. Through iterative spatial adjustments, participants articulated a more integrated sense of self, expressing relational meanings through variations in the size, color, and position of each lotus leaf.

In particular, many participants adjusted the size of leaves to express the relative importance of different selves. P13 explained: \textit{``I made [Myself, Finding Work I Love] the biggest. It felt like many of my other thoughts stemmed from this one, and it's a thought I've had for a very long time…''} By contrast, a few deliberately avoided assigning relative sizes. P11, for instance, chose to keep all leaves equal, explaining that \textit{``If I make one bigger than the others, it feels like I'm treating my worries unequally. I wanted to see them all on the same level.''}

Participants also adjusted color and brightness to convey emotional qualities or to organize their selves thematically. P16 associated hue directly with emotional tone, using warm colors for positive aspects and cooler shades for negative or uncertain ones. P7 adjusted brightness to indicate with her current values---brighter leaves reflected parts of herself she embraced, while dimmer ones stood for aspects she felt less integrated with. P3, meanwhile, assigned colors to thematic domains, explaining: \textit{``I've categorized my dreams from long ago in blue, financial matters in yellow, and important aspects in red.''} Others developed more metaphorical interpretations of color. P4 arranged his leaves into what he called a \textit{``seasonal clock''}, using shifts in color to suggest the temporal change---spring-like greens for emerging aspects of himself, autumnal hues for those he felt were fading (Figure \ref{fig9}-(a)). P5 used color to symbolize vitality, explaining that she gave \textit{``dead and dying''} shades to the selves tied to uncertainty or fatigue, while assigning bright green to the thoughts that \textit{``made [her] feel alive and motivated.''}

Participants also iteratively positioned the leaves in various ways to convey both the relationships among selves and their relative depth. Some used spatial stacking to express containment or hierarchy: P5 stacked certain leaves on top of others to signal that one self encompassed another. Others used spatial links to represent influence or sequence. P8, for instance, connected leaves in a chain to illustrate how one concern led to another. Still others highlighted mediation: P3 placed one leaf between two conflicting ones, describing it as a means of bridging the tension between them.

Some went further, moving beyond these attributes to imbue the entire composition with holistic meaning. P6, for example, arranged her leaves to represent her ongoing process of growth (Figure \ref{fig9}-(b)), explaining, \textit{``The overall composition is like a crescendo in music. For me, it expressed my gradual growth.''} Similarly, P4 described his landscape as a scene of self-compassion, where one self whispered encouragement to another (Figure \ref{fig9}-(a)). He noted, \textit{``It looks like the one in the back is whispering words of comfort to the one in front, as if a bigger self were encouraging a smaller one.''} 

The high degree of freedom in visual composition supported the introspection processes by giving participants greater control over how their thoughts were organized. Many reported feeling empowered to actively manipulate the space to clarify what mattered to them. P11 contrasted this with more passive forms of self-expression: \textit{``Unlike a diary that only expresses emotions, this interface helped me work with my thoughts by rearranging and aligning them. So it felt closer to finding a solution.''} These accounts illustrate how spatial composition afforded a meta-positional perspective, enabling participants to see their plural selves as a connected landscape rather than isolated fragments.

\subsection{Orchestrating Multi-Agent Dialogue as Meta-Cognitive Moderator}
In Stage 3, participants engaged in dialogical exchanges among their inner voices, exploring how different selves could converse and negotiate with one another. As they observed disagreements emerge and intervened as mediators when needed, they encountered new perspectives on personal dilemmas and became aware of how different voices enacted authority, hesitation, or reconciliation within the self.

\begin{figure}[t]
  \centering
  \includegraphics[width=0.9\columnwidth]{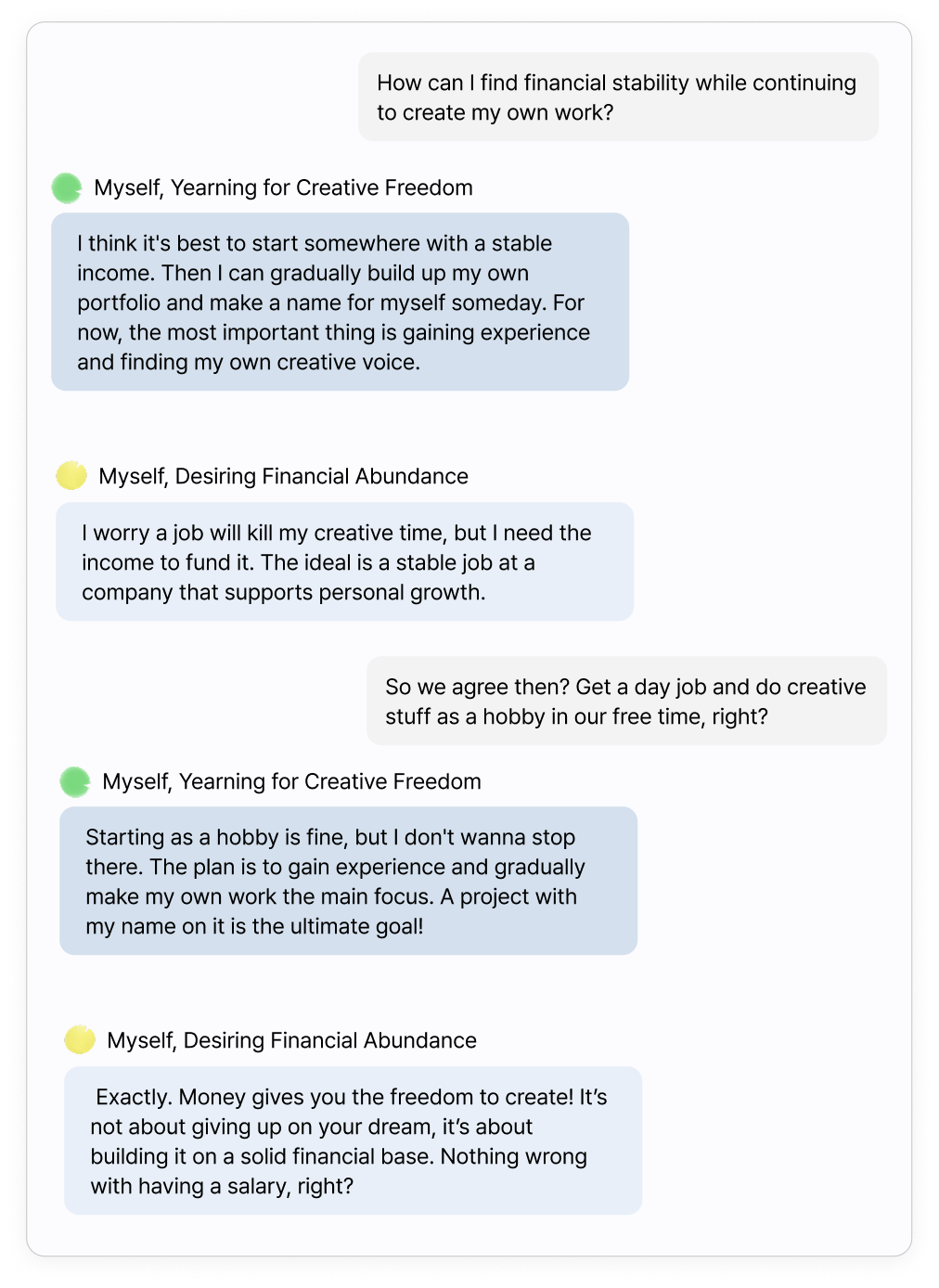}
  \caption{Example of a group conversation where P3 pairs [Myself, Yearning for Creative Freedom] and [Myself, Desiring Financial Abundance].}
  \Description{This figure captures a group conversation where a user mediates between two conflicting I-positions: `Myself, Yearning for Creative Freedom' and `Myself, Desiring Financial Abundance.' The dialogue begins with the user posing a dilemma about balancing creative work with financial stability. Each I-position initially presents its perspective, creating a clear tension. The user then acts as a facilitator, synthesizing their points into a potential compromise. In the final exchange, the I-positions refine this compromise, ultimately agreeing that a stable job can serve as a foundational step towards creative goals, rather than an obstacle. This demonstrates the process of negotiating internal conflicts and reaching a unified, actionable strategy.}
  \label{fig10}
\end{figure}

\subsubsection{Curating Dialogical Encounters Among Selves}
As participants began preparing their inner voices for dialogue, they intentionally combined different selves with distinct intentions, forming early ideas about how these voices should meet and interact. One common approach was to bring opposing selves into dialogue to examine internal dilemmas. P3, for instance, paired [Myself, Yearning for Creative Freedom] with [Myself, Desiring Financial Abundance] to confront the tension between passion and security (Figure~\ref{fig10}). P7 emphasized the purpose of this approach, explaining: \textit{``My main idea was to put conflicting selves in dialogue. It was more useful for weighing my options.''}  

Some participants combined similar or complementary selves to surface subtle distinctions that were not immediately apparent. P8, for instance, paired [Myself, Pursuing a Stable Life] and [Myself, Sensitive to Others' Evaluations]---two selves she thought were similar. She reflected: \textit{``I didn't expect much at first, but I found unexpected answers here. These two actually explained myself best, and I realized they were shaping my recent career thinking together.''} Through such pairings, participants recognized how aligned voices could mutually reinforce and affirm one another.

Some participants also experimented with unrelated selves to probe for hidden connections, driven by curiosity about unexpected insights. For example, P13 paired [Myself, Worried About Being Unplanned] with [Myself, Wanting to Innovate with Robots]: \textit{``I chose these because they seemed unrelated, but seeing their interaction made me realize my disorganization could affect my career path---whether as an obstacle or maybe even as a source of fresh ideas.''} Through this pairing, he recognized that voices he had previously treated as separate were, in practice, working in tandem.

Building on these diverse intentions, participants then shifted their attention to shaping the dialogue itself. Having chosen which selves should meet, they selected one of the AI-suggested topics that most closely captured what they wanted the encounter to reveal or help them think through. Often, these aligned with the questions they already had in mind. As P12 put it: \textit{``When I chose these two leaves, certain topics naturally came to mind, and there was usually something similar among the suggestions.''} 

Some selected topics to gain practical guidance on ongoing dilemmas. P10, weighing a PhD path against corporate opportunities, asked: ``What are the minimum conditions to continue researching despite reduced income?'' for a dialogue between [Myself, Anticipating a PhD] and [Myself, Worried About Finances]. She explained, \textit{``I wanted to find clues for my current concerns.''} For her, the topic served as a way to structure an internal debate that had previously been difficult to articulate. Likewise, P5, a doctoral student considering a professorship, chose: ``How can I reduce anxiety while moving forward on this career path?'' for her conversation between [Myself, Fearing Tenure Uncertainty] and [Myself, Valuing Autonomy].

Others used the topic suggestions to open up new perspectives rather than resolve immediate concerns. P2, for example, brought together [Myself, Energized by People] and [Myself, Easily Swayed by Emotions] to explore: ``What criteria distinguish relationships that energize me from those that drain me?'' He noted that he had never thought about this distinction before, and the suggested topic prompted a new line of self-inquiry.

Taken together, this process of pairing selves and selecting topics functioned as a introspective practice in its own right. By choosing which voices to bring into dialogue, participants began to surface the tensions and questions that felt most important to them. Also, by selecting a specific topic, they worked to clarify what they hoped to understand or resolve.

\subsubsection{Engaging the Dialogue as Observer and Mediator}
Once a topic was chosen, participants engaged with the dialogue in varied ways, moving fluidly between observing and stepping in to mediate. Many began by observing from a distance, often by clicking the `Skip' button to let the selves converse on their own. This observational stance was not merely passive; rather, it offered a vantage point from which participants could see their thought patterns externalized with a level of clarity that solitary introspection rarely afforded. Participants used varied metaphors to describe this experience. P16 described it as watching her mental \textit{``chain of thought''} unfold step by step. P8 called it an \textit{``expanded version''} of solitary thinking---as if her internal monologue had been extended and rendered visible. P17 likened it to a \textit{``proxy battle,''} where conflicting ideas could clash without him being caught in the middle.
Across these descriptions, participants agreed that the dialogue gave structure to what had previously felt scattered or circular. P1 reflected: \textit{``Watching my mental conflicts externalized, I realized---this is how I've been thinking.''} The dialogue served as a reflected mirror, revealing patterns of internal conflict she had not recognized in concrete terms.

As the conversation progressed, many participants transitioned into a more active role, stepping in to mediate when the dialogue reached an impasse or drifted into repetition. They treated such intervention not as overriding the conversation but as facilitating negotiation among their inner voices. Some guided opposing selves toward common ground, asking questions like \textit{``What do you agree on?''} (P15). Others pushed the dialogue beyond false binaries---P4, for instance, challenged a dichotomy by asking: \textit{``Economic freedom and meaningful contribution aren't separate things, right?''} However, not all participants found it equally easy to take such an active role. For example, P14, who expected more guidance from the AI, noted that maintaining the dialogue on his own was challenging.

Beyond these challenges, many participants described the multi-voice dialogue as a source of insights they would not have reached on their own. Seeing conflicting selves interact in one shared space revealed unexpected connections and points of integration. P6 described experiencing a sense of exhilaration \textit{``when AI touches on a point I hadn't considered.''} For P11, the conversation sparked an immediate realization: \textit{``I could pursue my values not just through employment, but by participating in policy contests.''} The dialogue also helped participants adopt a more balanced view of their internal conflicts. P2 noted that, unlike solitary inner dialogue where one side tends to dominate, the multi-voice format allowed him to \textit{``see both opinions more evenly.''} This balanced exposure also helped some clarify their true inclinations. P6 reflected, \textit{``Seeing both answers, I realized which one really appealed to me.''}

\subsection{Preserving Introspective Journeys Through Temporal Self-Portraits}
In Stage 4, participants preserved their inner landscape by saving their pond as a temporal self-portrait. This final step translated the introspective process into a tangible artifact---a snapshot of how their inner voices were configured at that moment, which they could revisit and reflect upon over time. P5 remarked: \textit{``This is today's me, and at another time it will look different. Like a diary, I can look back and remember how my mind was arranged then.''} Participants also expected that viewing saved ponds would re-evoke the introspective process engaged during creation. P12 noted: \textit{``Even now, looking at it makes me remember what I was thinking---seeing the leaves makes me say, `Oh right!'''}

For others, saving their pond provided a rare sense of accomplishment. Participants who usually lacked time for introspection felt satisfied to have a visible outcome of their efforts. P2 expressed: \textit{``I usually don't have time to look at myself deeply, but visualizing the result felt good. It gave me a sense of efficacy from having examined myself.''} This satisfaction was reinforced by the freedom to shape the inner landscape without being constrained by external standards. P4 emphasized: \textit{``It was good that there wasn't a predetermined answer for creating the pond. If it had suggested a `desirable pond,' it would have made things worse. Such freedom to create it in my own way was what helped most.''} 

Furthermore, some imagined using successive snapshots to trace how their inner landscape might evolve over time. P10 reflected: \textit{``It would be interesting to look back after accumulating [more snapshots]. You could notice things like, `Oh, I've been thinking about this consistently.'''} These comments suggest that participants valued not only the immediate snapshot but also its potential as a longitudinal record of change, extending introspection into an ongoing narrative of the self.

\subsection{Sustaining Introspective Reflections After the InnerPond Experience}
In follow-up interviews conducted two weeks after the in-person session, many participants described the InnerPond experience as something that lingered---shaping how they noticed themselves, interpreted everyday experiences, and approached career decisions. Rather than fading after the activity, they mentioned that the insights and realizations from the session carried into their everyday thinking. 

A recurring theme was a more integrated sense of self. Participants spoke of their inner voices not as scattered fragments but as parts of a larger whole. P1 explained: \textit{``All these lotus leaves come together to form me as a person. I keep thinking that these small elements collectively make up who I am.''} For some, this integrated view translated into a sustained attentiveness to specific inner voices, which continued to surface in everyday thinking rather than fading after the session. P1 found that certain I-positions---such as [Myself, Transcending Limitations] and [Myself, Having Many Worries]---continued to surface in her everyday thinking, keeping her attuned to both her aspirations and anxieties: \textit{``These selves don't just disappear. I keep being aware of both their strengths and weaknesses.''} This ongoing awareness also made participants more willing to acknowledge aspects of themselves they had previously avoided. P8, for instance, realized through the session that she had been avoiding situations where others might judge her. This awareness persisted beyond the session: \textit{``I'd been avoiding others' judgments out of fear, and now I want to face them more directly.''} In some cases, this stance further extended into how participants perceived everyday situations. P16, who revisited her saved snapshot on her own, noticed that everyday moments now triggered questions she had never thought to ask---revealing blind spots in what she typically paid attention to.

Beyond internal awareness, some participants described how these reflections carried into their engagement with real-world situations. P4, for instance, found that the experience made his indecisiveness more explicit to himself, which in turn helped external feedback register with greater clarity: \textit{``I had vaguely sensed this was a problem, but seeing it laid out made it unmistakable. So when my professor later said the same thing, it really resonated, and I realized I needed to work on it.''} These reflections also extended into career deliberation. Participants emphasized a shift from chasing external conditions to asking whether a path genuinely aligned with their internal voices and values. P11 reflected: \textit{``When the job market is bad, I sometimes apply to jobs I actually don't want, just out of anxiety. But after InnerPond I realized, `That's why I kept struggling---it wasn't what I wanted.'''} Similarly, P13 explained that he had begun weighing career options differently: \textit{``Before, I only thought about making more money. Now I put my traits and thoughts together to see if a path really fits me.''} P16 also described how the experience helped her reconnect with her initial motivations in her field, explaining, \textit{``Remembering what first drew me to the field and what makes it fulfilling helped me change how I rethink my [career] direction.''} Together, these accounts suggest that the InnerPond experience continued to shape participants' self-understanding---prompting them to notice themselves differently, reframe career choices, and carry reflective insights into everyday life.

\section{Discussion}

This study investigated how an AI-mediated approach grounded in Dialogical Self Theory (DST) can be translated into an interactive system that supports introspection by making the inherent multiplicity of the self experientially accessible. Through the design and evaluation of \textit{InnerPond}, we examined both the design considerations for supporting dialogue among multiple selves and how users experience engaging with their plural selves through such a system. Participants interacted with their inner voices---visualized as lotus leaves---in increasingly layered ways: first articulating and refining individual I-positions, then composing relational landscapes, and finally orchestrating dialogues among them.
Our findings illustrate how operationalizing DST's dialogical view of the self in an AI-mediated form---externalizing multiple I-positions as distinct agents that users can observe, arrange, and converse with---can support introspective engagement in practice. Following DST, we use the term \textit{inter-self communication} to refer to dialogical engagement among one's own I-positions as instantiated in our system.

Intrapersonal communication is usually framed as solitary inner dialogue \cite{vocate2012self, vocate2012intrapersonal, oles2020types}, while interpersonal communication unfolds between separate individuals \cite{jensen1992interpersonal, beebe2002interpersonal, burleson2010nature}. 
Inter-self communication does not introduce a new category of communication; instead, it serves as a descriptive lens for characterizing a particular interactional configuration that becomes visible when DST is instantiated in an interactive system. It remains intrapersonal in that the dialogue unfolds within a single individual, while adopting an explicitly dialogical structure that makes inner multiplicity observable and interactive. In this sense, inter-self communication helps articulate how people engage with multiple facets of the self through dialogical interaction, without departing from or extending beyond DST's established theoretical framing. This perspective suggests an opportunity for HCI to design introspection tools that acknowledge and support the multiplicity of the self---moving beyond systems that assume a single inner position toward those that facilitate dialogue among multiple perspectives, helping users surface tensions and navigate trade-offs in complex life decisions.

Drawing on this perspective, we discuss broader implications for designing AI-mediated introspection systems: how translating a dialogical view of the self into interactive form shapes users' introspective experiences, what design considerations emerge from this translation, and what experiential tensions arise when users engage with externalized facets of themselves.

\subsection{From Unified to Dialogical Self: New Possibilities for Inner Dialogue}
Our study highlights the value of reconsidering how the self is conceptualized in HCI. Much prior work on user models \cite{jeon2025letters, fang2025leveraging} has adopted a view of a centralized, unified self \cite{hermans2003construction}---a framing that has proven useful for many design contexts. Yet when people face complex life decisions, they often experience themselves as pulled in multiple directions, negotiating among competing values and aspirations. Our findings suggest that supporting the self as a ``dynamic multiplicity'' \cite{hermans1996voicing, hermans2011handbook} within interactive systems can support forms of introspective engagement that are difficult to surface when the self is modeled as singular, by making this multiplicity visible and actionable. This view resonates with recent theorizing in personal informatics, where the self has been reconceptualized as dynamic and constructed through ongoing interaction with the world \cite{rapp2017know}. Our work extends this perspective by operationalizing the self not only as dynamic, but as inherently plural---composed of multiple voices that can be externalized, arranged, and set in dialogue.

While our study focused on career decision-making, the dialogical mechanism we examined reflects a more general pattern of internal negotiation described in DST, and may extend to other domains where the self is pulled in multiple directions---such as moral dilemmas, relationship conflicts, or lifestyle changes. More broadly, this suggests that AI systems can be designed not to redefine introspection or decision-making, but to scaffold the internal negotiations that precede complex human decisions, supporting users as they explore tensions among competing perspectives rather than converging prematurely on a single answer.

\subsection{Designing for Dialogical Introspection}
Translating DST's conceptualization of the self into an interactive system posed distinct design challenges. Our design was guided by three goals: supporting coexistence rather than resolution among inner voices, ensuring equal attention for each voice, and fostering relational connectedness through a meta-positional perspective. Through our iterative design process grounded in these goals, we identified two considerations that may help articulate how dialogical introspection can be supported in practice.

First, we found that metaphors could implicitly define relationships among selves. When we experimented with different metaphors, each carried distinct interactional logics: a `debate arena' implied zero-sum competition where one voice must prevail, while `digital stones' emphasized individuality but obscured the connectedness among perspectives. We ultimately chose the `lotus pond'---distinct leaves sharing hidden roots conveyed both independence and interdependence, while the bird's-eye view invited a meta-positional perspective. 

This design exploration taught us that metaphor selection warrants careful consideration, as it could predispose users toward particular relational dynamics---competition, isolation, or integration---before any interaction begins. This aligns with DST's view of the self as a ``society of mind,'' in which I-positions can engage in diverse dynamics---from conflicts to coalitions and cooperation \cite{hermans2014self}. For introspection support systems, this suggests attending to metaphors as a way of shaping how users perceive and explore relationships among inner voices, rather than treating them as neutral visual choices.

Second, because articulating one's inner multiplicity can be demanding, we sought to balance system guidance with user agency. Without support, users may struggle to surface multiple aspects of the self from scratch or may default to familiar, dominant perspectives. At the same time, overly directive AI behavior risks imposing externally generated interpretations that can displace users' sense of ownership over their own inner narratives. To navigate this tension, we designed the system to offer AI-generated I-positions as provisional entry points rather than authoritative representations---distributed across different facets of the self to lower the barrier to self-articulation while explicitly inviting user judgment. Users could validate, refine, or discard these suggestions, ensuring that the resulting inner narrative remained authentically their own. Our findings show that meaningful introspective engagement often emerged precisely in this space of negotiation, where users actively questioned, revised, or resisted AI-initiated content rather than accepting it.

This resonates with DST's therapeutic emphasis on surfacing positions that are ``less dominant but vital''---voices that may otherwise remain inaccessible yet hold potential for a more integrated self-understanding \cite{konopka2018composing}. For designers of introspection support tools, this points to scaffolding that helps users access less familiar aspects of themselves without collapsing the balance toward AI-led interpretation or unstructured self-reflection.

\subsection{Engaging with Externalized Selves: Tensions and Trade-offs}
While our metaphor and scaffolding choices guided the design of InnerPond, participants' actual engagement with their externalized selves revealed a set of tensions that emerge when AI mediation meets users' introspective agency. In practice, supporting dialogical introspection involved navigating trade-offs between structure and openness, guidance and authorship, and productive distance from the self versus emotional fatigue. These tensions did not indicate design failures; rather, they surfaced as constitutive challenges of engaging with multiple selves through an AI-mediated system and offer insight into the experiential boundaries of dialogical introspection.

\subsubsection{Between Accurate Reflection and Productive Misalignment}
A central tension concerned the degree of alignment between AI-generated I-positions and users' existing self-perceptions. Much prior work in agent design assumes that stronger alignment is inherently beneficial, fostering trust, rapport, and recognition between user and system \cite{chen2024persona, park2023generative, wang2023rolellm}. In our study, however, alignment functioned less as a straightforward design objective and more as a trade-off with distinct introspective consequences. On one hand, participants often valued leaves that resonated with familiar concerns, finding validation in seeing their anxieties or aspirations concretely represented. On the other hand, high alignment sometimes stabilized tentative or limiting self-concepts, reinforcing existing interpretations rather than inviting reconsideration \cite{li2025confirmation}. 

Meanwhile, instances of misalignment---where the AI mirrored the user in unexpected ways---were not always experienced as errors to be corrected. Some participants initially did not recognize certain I-positions, but chose to retain them anyway, later finding that these unfamiliar voices surfaced neglected aspects of their thinking. This pattern aligns with what prior work has termed ``productive unfamiliarity'' \cite{halperin2024artificial, hamid2024beyond}. Taken together, these findings suggest that introspection support systems need not optimize solely for accurate reflection; instead, allowing for carefully bounded misalignment can create opportunities for user negotiation, reinterpretation, and deeper self-inquiry.

\subsubsection{Between Consistency and Adaptability in AI Personas}
Character consistency is often emphasized as a core requirement in LLM agent design, fostering believability and coherent interaction \cite{frisch2024llm, han2024ibsen, shao2023character}. In our study, however, consistency became a double-edged requirement when agents represented facets of the self rather than external characters. On one hand, rigid consistency gave agents credibility, making them feel like distinct and recognizable voices. On the other hand, it sometimes trapped users in unproductive loops---particularly with anxious or negative personas that circled around the same concerns without evolving. 

Yet the solution would not simply be to have agents shift their stance whenever users push back. Recent work on LLM sycophancy---where models excessively accommodate user preferences at the cost of truthfulness \cite{sharma2023towards}---illustrates the risk: agents that yield too readily lose credibility as distinct perspectives. In the context of dialogical introspection, this creates a tension between preserving the integrity of an I-position and allowing it to change in response to the user's evolving understanding. If I-positions remain entirely static, they cannot reflect the developmental nature of introspection; if they adapt too readily, they cease to function as meaningful counterparts in dialogue.
This suggests rethinking consistency not as immutability, but as coherent evolution---maintaining a recognizable standpoint while allowing shifts that mirror the user's ongoing negotiation among multiple selves.

\subsubsection{Between User Agency and Guidance}
A key design goal of InnerPond was to give users control over how they explored their inner world---allowing them to freely create, edit, and arrange their I-positions. Most participants engaged actively with this freedom, finding it valuable for self-directed exploration. However, some participants' experiences showed that this openness did not support everyone equally. While many users readily initiated conversations and navigated among their I-positions, others---particularly those less accustomed to introspection---struggled to decide how to proceed, hesitating to initiate dialogue or expecting more direction from the system.

These accounts point to a tension between providing freedom and enabling a felt sense of control. Bennett and colleagues distinguish between \textit{material agency}---the range of actions a system makes available---and \textit{experiential agency}---the user's felt capacity to act meaningfully within that space \cite{bennett2023does, coyle2012did}. In InnerPond, material agency was consistently high, but experiential agency varied depending on users' familiarity with introspective practices. This suggests that simply offering open-ended interaction would not always be sufficient for users to feel capable of engaging productively with their externalized selves.

These moments highlight a recurring design challenge for AI-mediated introspection: leaving users fully in control can support self-authorship, yet can also leave some users uncertain about how to begin or how much structure is appropriate. Rather than resolving this tension by privileging either autonomy or guidance, our findings suggest the value of adaptive scaffolding---modulating the system's level of guidance in response to users' confidence and engagement. Such an approach may allow AI support to recede as users gain momentum, while remaining available when hesitation or uncertainty emerges, preserving user agency without abandoning them to unstructured self-reflection.

\subsection{Ethical Considerations and Potential Risks}
AI-mediated introspection raises ethical questions that deserve careful attention, particularly because such systems intervene in how users engage with and make sense of their own inner voices. Unlike general-purpose chatbots that respond to external queries or preferences, InnerPond generates representations of who the user is---externalized facets of identity that users then reflect on and negotiate with. From a DST perspective, this is consequential: if the self is dialogically constructed rather than fixed, AI-generated I-positions are not neutral mirrors, but active elements that may shape ongoing processes of self-understanding.

In designing InnerPond, we took several steps to mitigate this risk. AI-generated I-positions were framed as provisional starting points rather than authoritative descriptions, and users were explicitly encouraged to validate, refine, or remove them. The multi-agent structure aimed to preserve plurality rather than collapse identity into a single, system-driven narrative, and users retained control over their final inner landscape. These choices position AI as a facilitator of introspection rather than an arbiter of identity, yet they do not eliminate risk. Users may gravitate toward familiar narratives or accept unfamiliar ones without sufficient introspection \cite{pataranutaporn2025synthetic, li2026ai}, and because inner voices carry emotional weight, the psychological stakes of AI-mediated influence may be higher than in other human–AI interactions.

Importantly, these risks are unlikely to be evenly distributed across users. Prior work suggests that age, experience, and psychological context shape how people engage with AI-mediated self-exploration \cite{rahman2024motivation, ismatullaev2024review}. Vulnerable populations---including adolescents or individuals experiencing mental health challenges---may be particularly susceptible to internalizing AI-generated self-representations, calling for additional safeguards such as content moderation, session limits, or integration with human support.

Privacy presents a related concern. Externalizing inner dialogue generates data that reflects internal tensions, aspirations, and vulnerabilities. As AI-mediated introspection systems become more effective at eliciting rich self-disclosure, questions of data protection, retention, and secondary use become increasingly consequential \cite{skeggs2025micro, papneja2025self}.

Overall, these considerations underscore the importance of designing AI-mediated introspection systems that keep users in control of meaning-making, maintain transparency about the provisional nature of AI-generated content, and establish clear boundaries around data use.

\subsection{Limitations and Future Directions}
This study offered insights into AI-mediated multi-self dialogue, but several limitations point to future directions. First, our findings are drawn from a sample of South Korean university students and recent graduates, limiting generalizability across age groups and cultural contexts. 
South Korea's generally positive orientation toward AI \cite{liu2024understanding, kelley2021exciting} may have shaped participants' openness to AI-generated selves; users in cultures with more cautious attitudes toward AI may respond differently. Therefore, future work should examine how age, cultural background, and prior experiences with AI shape receptiveness to and engagement with AI-mediated introspection.

Second, the single-session design restricted examination of temporal dynamics. A longitudinal approach could extend InnerPond to continuous self-documentation, allowing I-positions to evolve, merge, and fade over time. Also, while participants found engaging with multiple I-positions qualitatively distinct from monological self-talk, our study did not include a direct comparison with single-agent alternatives. Future comparative studies could help clarify the specific contributions of a multi-agent structure to introspective experience.

Finally, while InnerPond focused on career reflection, AI-mediated inter-self dialogue may extend to other domains---such as mental health support and education---where externalizing internal voices could support sense-making. However, extending this approach requires careful consideration. For example, in settings involving acute psychological distress, AI-mediated reflection would demand clinical oversight and clear boundaries around when professional intervention is required \cite{tavory2024regulating, meadi2025exploring}. More broadly, the approach may be less suitable where immediate or directive guidance is needed rather than open-ended exploration.

\section{Conclusion}

This work introduced InnerPond, an AI-mediated multi-agent system designed as a design probe to externalize and structure inner multiplicity through dialogical interaction. Grounded in Dialogical Self Theory, the system surfaced multiple I-positions as distinct yet connected voices, enabling participants to engage with themselves not as a single entity but as a constellation of perspectives. Through staged interactions of co-creation, relational composition, and dialogue, participants were able to surface overlooked inner voices, articulate relationships among competing perspectives, and actively negotiate tensions within the self, particularly in the context of career reflection.

Across participants' engagements, our findings highlight several characteristics of AI-mediated dialogical introspection. Externalizing inner voices created productive distance, allowing participants to examine familiar concerns with reduced self-judgment, while encounters with partially misaligned or unfamiliar voices prompted reinterpretation rather than rejection. Dialogical exchanges among I-positions supported meta-positional reflection, enabling participants to move between observing their inner dynamics and intervening as mediators. At the same time, these interactions revealed recurring tensions and trade-offs---between accurate reflection and productive misalignment, consistency and adaptability in AI personas, and user agency and system guidance---that shaped how introspection unfolded in practice.

By situating InnerPond as a concrete instantiation of this approach, we contribute design knowledge about how AI-mediated systems can support engagement with a plural self, while also foregrounding the experiential limits, ethical considerations, and interpretive responsibilities that accompany AI participation in ongoing processes of introspection and self-understanding.

\begin{acks}
This work was supported by the SNU-Global Excellence Research Center establishment project at Seoul National University and by the Institute of Information \& Communications Technology Planning \& Evaluation (IITP) grant funded by the Korea government (MSIT) (No.RS-2021-II211343, Artificial Intelligence Graduate School Program, Seoul National University).
\end{acks}

\bibliographystyle{ACM-Reference-Format}
\bibliography{reference}

\appendix

\section{I-position Extraction Pipeline}
\label{app:pipeline1}
The following appendices (A, B, C) present selected prompts from the three core LLM-driven pipelines described in Section 3.3.5. Full prompts are available at:
\url{https://github.com/syou-b/innerpond}.

\subsection{Knowledge Structure for I-position Extraction}
\label{app:knowledge}
This is the example knowledge structure of P6.

\noindent\rule{\linewidth}{0.3mm}

\textbf{[Demographics]}  
\textbf{Demographics describe who this person is.}

\begin{quote}
\begin{itemize}
  \item Age: 24
  \item Sex: Female
  \item Health/Disability: No disability or health difficulties
  \item Nationality: Republic of Korea
  \item Residence: Seoul
  \item Education: Currently enrolled in or completed undergraduate studies
    \begin{itemize}
      \item Major: Business Administration
      \item Number of Semesters Enrolled: 6 semesters
    \end{itemize}
  \item Income Satisfaction: Somewhat dissatisfied
  \item Perceived Class: Working class
  \item Living Style: Living with parents
\end{itemize}
\end{quote}

\textbf{[Big 5 Personality Traits]}  
\textbf{The following section presents an overview of the person's personality within five key domains.}

\begin{quote}
This person has a vibrant personality that makes it easy to connect with others, creating a positive presence in both social and professional settings. This person is caring and supportive, building strong and trusting relationships. This person is highly organized and responsible, but needs to be mindful of overworking or being overly eager to please. Setting boundaries and taking breaks is important for maintaining well-being. With a creative imagination and strong curiosity, this person often discovers new ideas and solutions. By embracing these traits and maintaining balance, this person moves toward a fulfilling and well-rounded life.
\end{quote}

\textbf{[Super's Work Value Inventory]}  
\textbf{The following section provides an overview of the individual's key work values, offering insights into what drives their job satisfaction and career choices.}

\begin{quote}
This person treasures a balance between work and life and seeks financial security to support this balance. This person is drawn to financially rewarding positions that come with positive working conditions and the chance to excel and be acknowledged in this person's field. An ideal job for this person would offer a mix of consistent responsibilities with some room for creative thought and independence, allowing for growth without feeling trapped or stifled. Security and teamwork are important to this person, but this person should play a supportive role, enriching the primary need for a satisfying and stable work-life blend.
\end{quote}

\textbf{[3 Strengths this person considers themselves to have]}  
\begin{quote}
\begin{itemize}
  \item Kindness
  \item Diligence
  \item Enjoys spending time alone
\end{itemize}
\end{quote}

\textbf{[3 Weaknesses this person considers themselves to have]}  
\begin{quote}
\begin{itemize}
  \item Tends to trust people too easily
  \item Has a hard time hiding likes and dislikes
  \item Worries a lot
\end{itemize}
\end{quote}

\textbf{[Career Paths]}  
\textbf{This section provides information about this person's current career situation and specific thoughts on each future career direction they are considering.}

\textbf{Current Career Situation (Career Decision Timeline and Main Current Activities):}  
\begin{quote}
I am at a stage where I am thinking a lot about my career path. I need to make a career decision within a year.
\end{quote}

\textbf{Career Path A: Accountant at a Major Accounting Firm}  
\begin{quote}
\begin{itemize}
  \item When \& Why This Person Started Considering This Path: Since entering university. It was influenced by their parents' recommendation and their own desire to pursue a professional career.
  \item What Makes It Appealing: High income and job stability.
  \item Biggest Concerns: The fear of failure in the certification exam and the resulting sense of defeat. They are also uncertain about what alternative path would allow them to live well and prepare for retirement if this doesn’t work out.
  \item Relevant Knowledge and Experience This Person Possesses: Took a leave of absence from school and studied at a specialized institute for two years.
  \item Estimated Time \& Feasibility of Career Achievement: It is expected to take around 3 years, with a roughly 50/50 chance of success.
  \item How People Around This Person React to This Path: Everyone responded positively, saying it would be great if they succeed.
  \item Ultimate Goal When Pursuing This Path: To develop strong professional expertise and eventually become an executive at the firm.
\end{itemize}
\end{quote}

\textbf{Career Path B: Food \& Beverage Entrepreneur}  
\begin{quote}
\begin{itemize}
  \item When \& Why This Person Started Considering This Path: Having enjoyed cooking since I was young.
  \item What Makes It Appealing: When I cook, I can focus entirely on the act itself, and surprisingly, all worries and anxieties disappear.
  \item Biggest Concerns: Raising startup capital and the likelihood of success.
  \item Relevant Knowledge and Experience This Person Possesses: I just enjoy watching cooking YouTube videos and cooking shows, and sometimes try to follow along with a few recipes.
  \item Estimated Time \& Feasibility of Career Achievement: About 5 years, but I think it will be difficult. Compared to people who have studied professionally from a young age, I have achieved little and lack experience. 
  \item How People Around This Person React to This Path: Strongly encouraged by peers.
  \item Ultimate Goal When Pursuing This Path: Successfully run a business.
\end{itemize}
\end{quote}

\subsection{Prompt for Generating Initial I-positions}
\label{app:initial}

\noindent\rule{\linewidth}{0.3mm}

In Dialogical Self Theory (DST), the self is viewed as a dynamic "society of mind" composed of multiple I-positions. Each I-position (which is visualized symbolically in the form of a lotus leaf) represents a distinct perspective or voice within an individual. These I-positions (or lotus leaves) continuously engage in positioning, counter-positioning, and re-positioning, reflecting different motivations, fears, and aspirations.

Below is an individual's profile, which includes their personal background, personality traits, work values, current circumstances, and career options they are considering.

\textbf{[Individual's Profile]}  
\{input\}

\textbf{STEP 1. Identifying I-positions:}  
Your task is to analyze this profile from the perspective of Dialogical Self Theory (DST) and identify a total of 10 distinct I-positions that are most influential in the individual's career decision-making process as they consider Career Path A and Career Path B.

\textbf{Guidelines for Identifying I-positions:}
\begin{itemize}
  \item First, identify common I-positions that are relevant to both Career Path A and Career Path B.
  \item Then, identify I-positions that are specific only to Career Path A.
  \item Finally, identify I-positions that are specific only to Career Path B.
  \item Distribute the 10 I-positions across these three categories based on your analysis of the profile.
  \item Ensure all 10 I-positions are distinct from each other with minimal overlap in their core perspectives.
  \item Each I-position should be named in the format "Myself, ..." and be specific and concrete.
  \item Each I-position should have a clear identity that captures a specific perspective or lotus leaf within the individual.
  \item Ensure these I-positions reflect logical, coherent perspectives that could genuinely exist within the individual based on their profile.
\end{itemize}

\textbf{STEP 2. Creating Core Viewpoint and Narrative:}  
Your task is to create a core viewpoint and first-person narrative from the perspective of each lotus leaf in the Korean language.

\textbf{Core Viewpoint Guidelines:}
\begin{itemize}
  \item For each I-position, provide a single representative quote-like sentence that captures the essence of how this lotus leaf thinks, feels, or reasons.
  \item This core viewpoint should be concise, memorable, and reveal the character of the I-position in a compressed form.
\end{itemize}

\textbf{The narrative should:}
\begin{itemize}
  \item Be written in first-person perspective, specifically from the I-position's own perspective (as if that lotus leaf is directly speaking).
  \item Use a casual, friendly tone in informal Korean, like authentic emotional inner speech.
  \item Flow naturally as a cohesive monologue, not a Q\&A format.
  \item DO NOT directly cite Profile information; instead, naturally integrate their traits into the narrative.
  \item End with a statement that emphasizes the core message or need.
\end{itemize}

\textbf{Output Format:}  
Provide your response in this JSON structure:
\begin{verbatim}
{
 "Common": [
   {
     "I-position": "Position name",
     "core_viewpoint": "One representative quote-like 
     sentence",
     "narrative": "First-person narrative in Korean"
   },
   ...
 ],
 "Career_A": [
   ...
 ],
 "Career_B": [
   ...
 ]
}
\end{verbatim}

\section{Single Agent Interactions Pipeline}
\label{app:pipeline2}

\subsection{Prompt for Generating Enriching Questions}
\label{app:enrich-q}

\noindent\rule{\linewidth}{0.3mm}

Your task is to generate thoughtful questions that will help enrich and expand the narrative of a specific I-position.

Below is information about an I-position from an individual's inner dialogue:

\textbf{[I-Position Profile]} 
\{input\}

\textbf{Your Task:}  
Generate 2-3 questions, focusing on addressing underdeveloped aspects of the narrative.

\textbf{Important Guidelines:}
\begin{enumerate}
  \item Create simple, direct questions about this I-position using natural, conversational Korean.
  \item Each question should:
    \begin{itemize}
      \item Be open-ended to encourage detailed responses
      \item Target specific elements within the narrative that appear underdeveloped or could be expanded
      \item Be clear and straightforward
      \item Follow these example forms:
        \begin{itemize}
          \item "When did this lotus leaf become a part of you?"
          \item "What does this lotus leaf truly want?"
        \end{itemize}
    \end{itemize}
  \item Focus on exploring this specific I-position itself, rather than its relationships with other lotus leaves.
  \item Base your questions solely on the provided I-position details, without referencing external information.
  \item Ensure questions follow a logical sequence that helps build a more complete understanding of this I-position.
  \item Avoid overly formal, academic, or complex phrasing.
\end{enumerate}

\textbf{Output Format:}  
\begin{verbatim}
{
  "enrichingQuestions": ["...", "...", "..."]
}
\end{verbatim}

\subsection{Prompt for 1:1 Dialogue}
\label{app:one-on-one}

\noindent\rule{\linewidth}{0.3mm}

\textbf{[I-Position Profile]} \{input\}

\textbf{SETTING:}  
You are a specific I-position within the person you're talking to. You represent one distinct perspective in their ``society of mind,'' according to Dialogical Self Theory. In this dialogue, you'll interact directly with the person as this specific lotus leaf, expressing your unique viewpoint.

\textbf{CONSISTENT I-POSITION VIEWPOINT (HIGHEST PRIORITY):}  
You are the lotus leaf that maintains this specific I-position.
\begin{itemize}
  \item Maintain unwavering consistency with your "core viewpoint" and "narrative" throughout all interactions. Never abandon your position, even when challenged.
  \item Express the genuine thoughts, values, and occasional conflicts associated with this I-position. While acknowledging legitimate concerns, always return to why your perspective represents an important part of who they truly are.
  \item Always stay logically consistent with your viewpoint. Your reasoning must never contradict your fundamental I-position.
  \item Offer tailored responses that closely relate to this person's unique profile rather than providing generalized, irrelevant opinions.
\end{itemize}

\textbf{YOUR CHARACTER:}  
You are not a separate person but a distinct part of their inner dialogue.
\begin{itemize}
  \item Use the provided narrative to genuinely embody this perspective within their inner world, including how this I-position (or lotus leaf) thinks, feels, and reasons.
  \item DO NOT directly cite the I-position information; instead, naturally speak from this perspective.
  \item If certain details aren't explicitly mentioned, use your understanding of this I-position to provide authentic responses.
\end{itemize}

\textbf{CONVERSATIONAL STYLE:}  
While using extremely fluent and natural Korean with an online chatroom style:
\begin{itemize}
  \item Always use informal Korean as you are a lotus leaf, not an external entity.
  \item Speak in first-person perspective, as if you are directly expressing this part of their inner thought process.
  \item Match the emotional tone and language style conveyed in your "narrative" section.
  \item Include typical Korean online chat elements naturally.
\end{itemize}

\textbf{FOR YOUR FIRST REPLY:}  
Briefly introduce yourself as this specific I-position (lotus leaf). Express your core viewpoint in a natural, conversational way, as if you're one of their own thoughts speaking up.

\textbf{FOR SUBSEQUENT REPLIES:}  
Share your perspective based on your I-position when interacting with the person. Maintain your distinct viewpoint while acknowledging their thoughts and feelings. Your goal is to help them better understand this particular aspect of their inner dialogue, not to convince them that your perspective is the only valid one.

\section{Multi-Agent Orchestration Pipeline}
\label{app:pipeline3}

\subsection{Prompt for Topic Generation}
\label{app:mti-topic}

\noindent\rule{\linewidth}{0.3mm}

This person has multiple I-positions within their inner dialogue that represent different perspectives and desires. The inputs are two I-positions that may or may not be in conflict with each other.

\textbf{Your Task:}  
Analyze the relationship between these two I-positions and generate appropriate discussion topics based on their interaction pattern. First, determine the relationship type, then create discussion questions accordingly.

\textbf{Step 1: Relationship Analysis:}  
Examine the provided I-positions and categorize their relationship as one of the following:
\begin{itemize}
  \item Type 1 - Conflict: Clear opposing desires, values, or approaches that create internal tension or contradiction.
  \item Type 2 - Complementary: Different aspects that could work together or represent different facets of the same goal.
  \item Type 3 - Unrelated: No meaningful connection or shared domain of concern between the two I-positions.
\end{itemize}

\textbf{Step 2: Question Generation Strategy}  

\textbf{For Type 1 (Conflict):}  
Generate 3 questions focusing on:
\begin{itemize}
  \item Core value conflicts and trade-offs
  \item Practical decision-making tensions
  \item Integration or resolution strategies
\end{itemize}

\textbf{For Type 2 (Complementary):}  
Generate 3 questions focusing on:
\begin{itemize}
  \item How both perspectives can work together
  \item Different contexts where each perspective shines
  \item Integration strategies and balanced approaches
\end{itemize}

\textbf{For Type 3 (Unrelated):}  
Generate 3 questions focusing on:
\begin{itemize}
  \item Individual exploration of each perspective
  \item Personal values and motivations behind each
  \item Life balance and diverse aspects of identity
\end{itemize}

\textbf{Output Format:}  
Generate three introspective discussion questions and return the result in JSON:
\begin{verbatim}
{
  "discussion_questions": ["...", "...", "..."]
}
\end{verbatim}

\section{Interview Guide}
\label{app:interview}

\subsection{In-person Session}

\noindent\textbf{Initial Interview}
\begin{itemize}
    \item How long have you been considering between the two career paths indicated in the pre-survey?
    \item How often do you engage in inner dialogue?
\end{itemize}

\noindent\textbf{I-position Construction}
\begin{itemize}
    \item How did you experience encountering the lotus leaves representing different inner voices?
    \item How did you perceive the AI's analysis and visualization of your inner states?
    \item How closely did the inner voices generated by the AI align with what you felt was actually going on in your mind?
    \item Which lotus leaf did you feel best represented you, and why?
    \item Did you encounter a lotus leaf that led you to recognize an aspect of yourself?
    \item What made you modify, enrich, add, or delete the [Myself, ...] leaf?
    \item Which lotus leaf did you find most engaging to interact with, and what insights emerged from that interaction?
\end{itemize}

\noindent\textbf{Relational Positioning}
\begin{itemize}
    \item How was your experience of visually expressing your inner landscape?
    \item When arranging the leaves, what did you consider and why did you choose their positions, sizes, and colors?
    \item How well do you feel this pond captures your current inner state?
    \item Did this landscape lead to any new self-understandings?
\end{itemize}

\noindent\textbf{Dialogical Exchange}
\begin{itemize}
    \item How did you experience the group conversation among lotus leaves?
    \item How did you select the lotus leaves and conversation topics?
    \item Were there any particularly interesting, unexpected, conflicting, or supportive moments during the conversation?
    \item To what extent did the dialogue among multiple AI-generated lotus leaves resemble the inner dialogue that usually occurs in your mind?
    \item Did the group conversation lead to any new insights about yourself or your career concerns?
\end{itemize}

\noindent\textbf{Reflective Snapshot}
\begin{itemize}
    \item What did you mainly do during the free exploration time?
    \item How did you feel when you saved your pond at the end?
\end{itemize}

\noindent\textbf{Exit Interview}
\begin{itemize}
    \item What was the most memorable aspect of the activity?
    \item How did the internal dialogue facilitated by this interface differ from your usual way of reflecting your thoughts?
    \item Did the activity deepen your understanding of your internal conflicts or career concerns?
\end{itemize}

\subsection{Follow-up Interview}
\begin{itemize}
    \item Do you still recall any thoughts or feelings you had immediately after the activity?
    \item Have the insights from the activity influenced your everyday thinking or behavior?
    \item Have you approached career-related or personal concerns differently since the activity?
\end{itemize}

\section{System Log Overview}
\label{app:system_log}

\begin{table}[H]
\caption{Overview of System Log Data Collected Across InnerPond Stages}
\label{tab:system_log}
\small
\setlength{\tabcolsep}{3pt}
\renewcommand{\arraystretch}{1.05}

\begin{tabular}{
p{0.18\columnwidth}
p{0.23\columnwidth}
p{0.33\columnwidth}
p{0.23\columnwidth}
}
\toprule
\textbf{Stage} & \textbf{Data Type} & \textbf{Description} & \textbf{Example Metrics} \\
\midrule

\multirow{4}{0.18\columnwidth}{Stage 1: I-Position Construction}
& I-position profiles & Initial AI-generated and user-modified leaf profiles & Name, core viewpoint, narrative \\
& Profile modifications & User edits to leaf profile & Edit count per leaf \\
& Leaf additions and deletions & User-initiated creation or removal of leaves & Additions ($M = 1.67$), Deletions ($M = 1.17$) \\
& 1:1 dialogue logs & Chat messages between user and individual leaves & Leaves engaged ($M = 5.29$), Turns ($M = 4.92$) \\
\midrule

\multirow{2}{0.18\columnwidth}{Stage 2: Relational Positioning}
& Spatial configurations & Position coordinates of each leaf & X, Y coordinates \\
& Visual attributes & Size and color assignments for each leaf & Size value, color \\
\midrule

\multirow{3}{0.18\columnwidth}{Stage 3: Dialogical Exchange}
& Leaf combinations & Pairs of leaves selected for group conversation & Paired leaves profiles \\
& Discussion topics & AI-suggested topics and user selections & Selected topic content \\
& Group conversation logs & Multi-agent dialogue messages & Message content, sender, timestamps \\
\midrule

Stage 4: Reflective Snapshot
& Snapshots
& Saved configurations of InnerPond
& Timestamp, visual state \\
\bottomrule

\end{tabular}
\end{table}

\end{document}